\documentclass[journal]{IEEEtran}
\usepackage{amsfonts}
\IEEEoverridecommandlockouts

\ifCLASSINFOpdf
\else
\fi
\usepackage{multirow}
\usepackage{cite}
\usepackage{stfloats}
\usepackage{color}
\usepackage{epsfig}
\usepackage{graphicx}
\usepackage{subfigure}
\usepackage{psfig}
\usepackage{epsf}
\usepackage{slashbox}
\usepackage{float}
\usepackage{amssymb }
\usepackage[cmex10]{amsmath}
\usepackage{verbatim}
\usepackage{amsfonts}

\newtheorem{remark}{Remark}

\newtheorem{proof}{Proof}
\begin{document}

\title{System Outage Probability of PS-SWIPT Enabled Two-Way AF Relaying with Hardware Impairments}
\author{\IEEEauthorblockN{Zhipeng Liu,  Guangyue Lu, Yinghui Ye, and Xiaoli Chu,~\IEEEmembership{Senior Member,~IEEE}} 
\thanks{ Zhipeng Liu (zhipeng\_liu\_steve@163.com), Guangyue Lu (tonylugy@163.com) and  Yinghui Ye (connectyyh@126.com) are with the Shaanxi Key Laboratory of Information Communication Network and Security,  Xi'an University of Posts \& Telecommunications, China.}
 \thanks{Xiaoli Chu  (x.chu@sheffield.ac.uk) is with the Department of Electronic and Electrical Engineering, The University of Sheffield,  U.K.}
\thanks{This work was supported by the Postgraduate Innovation Fund of Xi'an University of Posts \& Telecommunications (CXJJLZ2019026), the Science and Technology Innovation Team of Shaanxi Province for Broadband Wireless and Application (2017KCT-30-02).}
}

\maketitle
\begin{abstract}
In this paper, we investigate the system outage probability of a simultaneous wireless information and power transfer (SWIPT) based two-way amplify-and-forward (AF) relay network, where transceiver hardware impairments (HIs) are considered. The energy-constrained relay node processes the received signals based on a power splitting protocol and the two terminals employ a selection combining (SC) scheme to exploit the signals from the direct and relaying links. Assuming independent but non-identically distributed Nakagami-$m$ fading channels,  we derive the system outage probability in a closed-form, which enables us to identify  two crucial ceiling effects on the system outage probability caused by HIs in   the high data rate regions, i.e., relay cooperation ceiling (RCC) and overall system ceiling (OSC). Specifically, the RCC prevents the relaying link from participating in cooperative communications, while the OSC  leaves the overall system in outage. Furthermore, we derive the  achievable diversity gain of the considered network, which shows that the diversity gain equals either the shape parameter of the direct link or zero.
Computer simulations are provided to validate the correctness of our analytical results, and study the effects of various system parameters on the system outage performance and the optimal power splitting ratio, as well as the energy efficiency. 
\end{abstract}

\section{Introduction}
\IEEEPARstart{S}{imultaneous} wireless information and power transfer (SWIPT) has been incorporated into relay networks to prolong the operation time of energy-constrained relay nodes via a time switching (TS) or power splitting (PS) scheme \cite{8214104,7953588,6552841}. Due to the potential in improving spectrum efficiency, SWIPT enabled two-way relaying, which allows two terminals to exchange their information through a relay, has been widely used in wireless sensor networks \cite{7120018}. For SWIPT enabled two-way relay networks (TWRNs), two popular protocols have been widely adopted by the existing works \cite{5605920, 8556567,7831382,7807356,7858707,7971948,7037438,8377371,8613860,8633928,8576644}. The multiple access broadcast (MABC) protocol requires only two time slots to accomplish information interchange between both terminals, but the direct link between the two terminals can not be utilized due to the half-duplex constraint. The time division broadcast (TDBC) protocol makes use of the direct link, but the information interchange needs three time slots \cite{5605920}.

\subsubsection{Related Works on SWIPT Enabled TWRNs with MABC} In \cite{8556567}, the authors derived the exact terminal-to-terminal (T2T) outage probability, system outage probability and system ergodic capacity for a PS-SWIPT enabled two-way amplify-and-forward (AF) relay system. For a TS-SWIPT two-way AF relay network, the closed-form system outage probability, ergodic sum-rate and sum symbol error rate were derived in \cite{7831382}. Song \emph{et al.} considered a SWIPT based two-way decode-and-forward (DF) relay network in \cite{7807356}, in which the PS/TS and time allocation ratios were jointly optimized to minimize the system outage probability. The sum-rate was maximized by optimizing the PS ratio in SWIPT based two-way DF relaying \cite{7858707}.
In \cite{7971948}, the authors proposed a relay selection strategy for a PS-SWIPT based two-way multiple relay network.

\subsubsection{Related Works on SWIPT Enabled TWRNs with TDBC} The T2T outage performance for three wireless power transfer policies in TS-SWIPT enabled two-way AF relay networks was analyzed in \cite{7037438}. In \cite{8377371}, the authors investigated the T2T performance for a TS-SWIPT enabled two-way relay system, where the relay node follows a hybrid-decode-amplify-forward (HDAF) strategy. The system outage performance of a PS-SWIPT based two-way DF relay system was studied in \cite{8613860,8633928}. For a TS-SWIPT based two-way DF relay system, the authors of \cite{8576644} proposed an optimal combining scheme at the relay to minimize the system outage probability.

We note that most of the existing works on SWIPT enabled TWRNs have assumed ideal hardware at all nodes \cite{7807356,7971948,7831382,7858707,8556567,8613860,8377371,8576644,8633928,7037438}. In practical systems, the radio frequency (RF) transceivers are afflicted with hardware impairments (HIs), such as I/Q imbalance, phase noise, high power amplifier (HPA) nonlinearities, etc \cite{6214150,990908,871400}. Several sophisticated algorithms have been developed to mitigate the effects of HIs, but the residual impairments remain and cause a substantial performance loss to the TWRNs in the medium and high data rate regions \cite{6891254,6630485}.
The authors in \cite{7247469} derived the T2T outage probability for TS-SWIPT based two-way cognitive relay networks (TWCRNs) with MABC/TDBC. Afterwards, this work was extended to PS-SWIPT based TWCRNs \cite{7835665}. It was shown that the HIs deteriorate substantially the T2T outage performance. For a TS-SWIPT based two-way HDAF relay system with TDBC, Solanki \emph{et al.} \cite{8851300} derived the T2T outage probability in the presence of HIs, and studied the impacts of HIs as well as various network parameters on the T2T outage performance. 
 The T2T outage probability considers only the outage event of one terminal, while the system outage probability jointly considers the outage evens of both terminals, as well as the correlation between the two links. Accordingly, quantifying the system outage probability is critical and is much more challenging than the T2T outage probability. However, the system outage probability of SWIPT enabled TWRNs under HIs has not been studied by any of the existing work.

In this paper, we study the system outage probability for a PS-SWIPT based two-way AF relay network, while considering the hardware impairments of all transceivers and the existence of a direct link between the two terminals. 
Different from MABC, TDBC leverages the direct link and has a low operational complexity at the relay node \cite{8613860}, thus this work considers TDBC instead of MABC.
All the channels follow independent but non-identically distributed Nakagami-$m$ fading. Our main contributions are listed below.
\begin{itemize}
  \item Based on the instantaneous end-to-end signal-to-noise-plus-distortion ratios (SNDRs), we derive a closed-form expression for the system outage probability  to characterize the influence of HIs.
  \item Based on the derived system outage probability, we identify two crucial ceiling effects caused by HIs, viz., relay cooperation ceiling (RCC) and overall system ceiling (OSC). More specifically, when the data rate is larger than the RCC threshold, the AF relaying operation ceases; when the data rate  goes beyond the OSC threshold that is larger than the RCC threshold,  the system goes in outage. 
      For a given transmission rate requirement, we obtain the maximum  tolerable level of HIs for maintaining an acceptable system outage probability.
  \item We derive the diversity gain for the considered network. It shows that when the date rate is larger than the OSC threshold, the diversity gain is zero; otherwise, the diversity gain equals the shape parameter of the direct link. This indicates  that HIs would prevent  the relaying link from contributing to the diversity gain, and that the diversity gain of the considered network with HIs is only determined by the direct link.


\end{itemize}

The rest of this paper is summarized as follows. In Section II, we first describe the system and channel models, and then obtain the end-to-end SNDRs. Section III derives a closed-form expression for the system outage probability to quantify the impact of HIs on system outage performance, and thereafter analyze the diversity gain of the considered network. In Section IV, simulation results are provided to validate the derived expressions and obtain some insights about system outage performance. Finally, Section V concludes the paper.

\emph{Notation:} $x\sim {\rm{Naka}}(\phi , \varphi)$ denotes a Nakagami random variable $x$ with fading severity parameter $\phi$ and average power $\varphi$. ${y} \sim \mathcal{CN}\left( {\vartheta,\psi} \right)$ represents a Gaussian random variable $y$ with mean $\vartheta$ and variance $\psi$. $\Pr \left(  \cdot  \right)$ and $\left|  \cdot  \right|$ denote the probability of an event and the absolute value of a number, respectively. $f_H(h)$ and $F_H(h)$ denote the probability density function (PDF) and the cumulative distribution function (CDF) of a random variable $H$, respectively. $\Gamma(n)=(n-1)!$ is the complete gamma function. $\gamma \left( {n,z} \right) = \int_0^z {{u^{n - 1}}{e^{ - u}}du}$ and $\Gamma \left( {n,z} \right) = \int_z^\infty {{u^{n - 1}}{e^{ - u}}du}$ denote the lower and upper incomplete gamma functions, respectively. ${K_n}(z)$ denotes the $n$-th order modified Bessel function of the second kind.

\section{System Model and Working Flow}

\subsection{System Model and Channel Model}
As illustrated in Fig. 1, we consider a SWIPT based two-way AF relay network with TDBC, where the energy-constrained relay $R$ harvests energy from the RF signal via a PS protocol and adopts the \lq\lq harvest-then-forward\rq\rq\;scheme to assist the information interchange between the two terminals $S_a$ and $S_b$ \cite{8613860,8633928}. A direct link exists between the two terminals. All involved nodes are equipped with a single antenna and work in the half-duplex mode.
Each transmission block $T$ is divided into three phases \cite{8361446,7279107}: two broadcast (BC) phases ($2T/3$) and one relay (RL) phase ($T/3$), as shown in Fig. 2.
In the first BC phase, terminal $S_a$ broadcasts its own signal $x_a$ to relay $R$ and terminal $S_b$. In the second BC phase, terminal $S_b$ transmits its signal $x_b$ to relay $R$ and terminal $S_a$.
After receiving the signal from $S_i$ ($i\!=\!a$ or $b$), relay $R$ splits the received power into two parts according to a PS ratio, i.e., a $\beta$ $\left(\beta \in (0,1)\right)$ portion of the received power for energy harvesting (EH) and the rest for information processing (IP).
In the RL phase, relay $R$ amplifies the received signals and broadcasts them to the two terminals $S_a$ and $S_b$ using all the harvested energy. 
\begin{figure}[!t]
  \centering
  \includegraphics[width=0.37\textwidth]{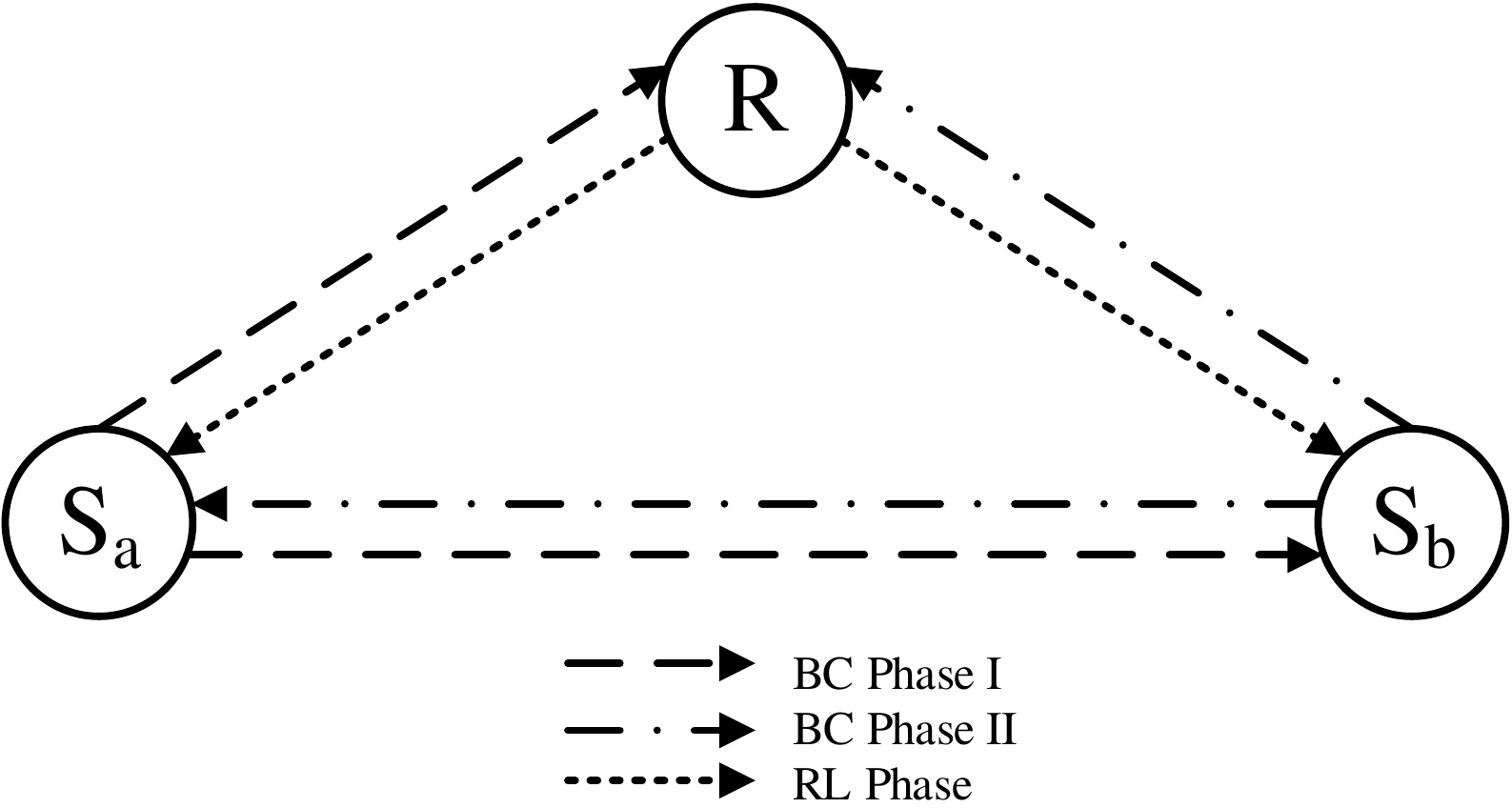}\\
  \caption{System model of PS-SWIPT based two-way AF relay network with TDBC.}
\end{figure}
\begin{figure}[!t]
  \centering
  \includegraphics[width=0.37\textwidth]{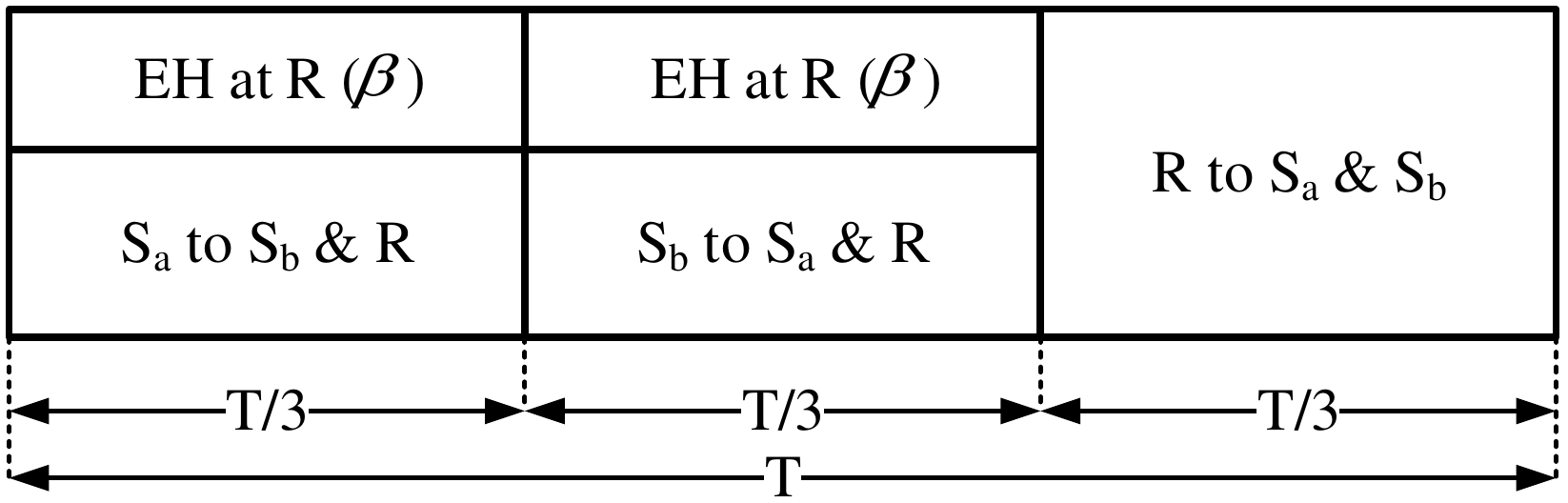}\\
  \caption{Transmission block structure for TDBC protocol.}
\end{figure}
\begin{figure}[!t]
  \centering
  \includegraphics[width=0.492\textwidth]{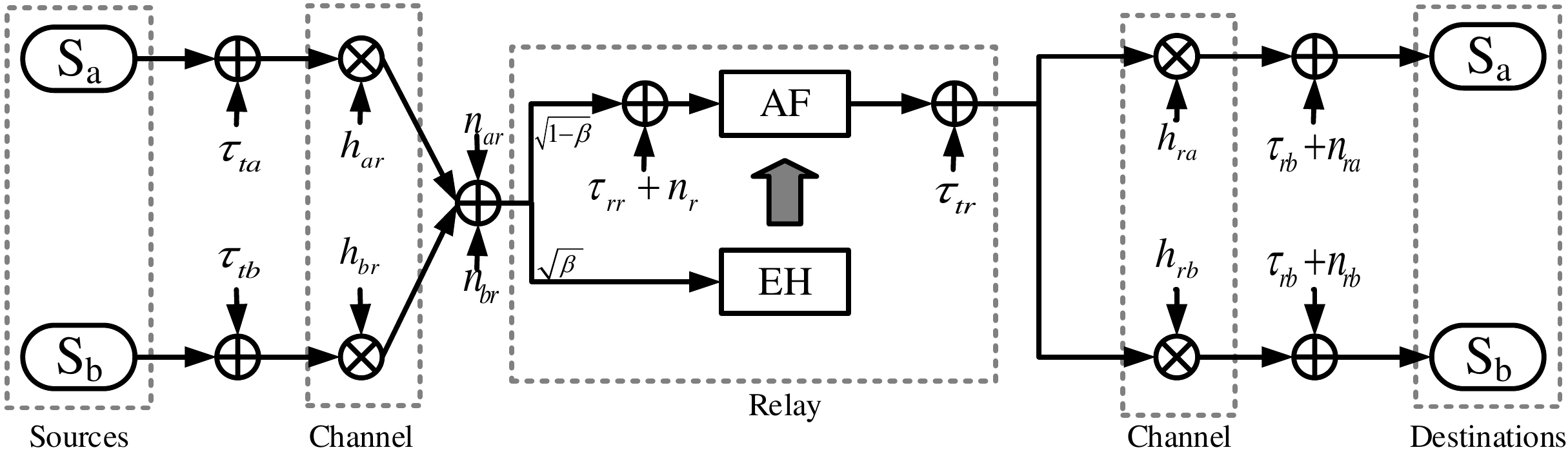}\\
  \caption{Block diagram of relaying link with HIs.}
\end{figure}


All the channels are quasi-static, reciprocal and subject to independent but non-identically distributed Nakagami-$m$ fading.
Specifically, the channel fading coefficient  between $S_i$ and $S_j$ is denoted by $h_{ij}\sim {\rm{Naka}}(m_d , \Omega_d)$, with $i,j \in \left\{ {a,b} \right\},\;i \ne j$. The channel fading coefficient of the $S_i$ to $R$ link is denoted by $h_{ir}\sim {\rm{Naka}}(m_i , \Omega_i)$. Since $h_{ij}$ and $h_{ir}$ follow the Nakagami distribution, the corresponding channel gains $|{h_{ij}}{|^2}$ and $|{h_{ir}}{|^2}$ follow the gamma distribution \cite{4376057}, and their PDF and  CDF are expressed as
\begin{align}\label{2}
  {f_V}\left( v \right) = \frac{1}{{\Gamma \left( {{m}} \right)\theta^{{m}}}}{v^{{m} - 1}}{e^{ - \frac{v}{{{\theta }}}}},
\end{align}
\begin{align}\label{25}
{F_V}\left( v \right) = \frac{1}{{\Gamma \left( {{m}} \right)}}\gamma \left( {{m},\;\frac{v}{{{\theta}}}} \right),
\end{align}
where $V \in \left\{ {|{h_{ij}}{|^2},|{h_{ir}}{|^2}} \right\}$, $m$ and $\theta$ denote the shape parameter and the scale parameter of the variable $V$, respectively. Note that when $V{\rm{ = }}|{h_{ij}}{|^2}$ $\left( {|{h_{ir}}{|^2}} \right)$, $m = {m_d}$ $\left( {{m_i}} \right)$ and $\theta {\rm{ = }}{\Omega _d}/{m_d}$ $\left( {{\Omega _i}/{m_i}} \right)$ respectively.

\subsection{TDBC Protocol}
In the first and second BC phases, $S_a$ and $S_b$ broadcast the information signal $x_a$ and $x_b$, respectively. 
For the direct link, the received signal at $S_j$ from $S_i$ ($i,j \in \left\{ {a,b} \right\},i \ne j$), under the presence of HIs, is expressed as
\begin{align}\label{3}
{y_{ij}} = {h_{ij}}\sqrt {{P_i}} \left( {{x_i} + {\tau _{ti}}} \right) + {\tau _{rj}} + {{n_{ij}}},
\end{align}
 where ${x_i}$ is the unit-power signal transmitted by $S_i$; $P_i$ denotes the transmit power of $S_i$;  ${\tau _{ti}} \sim \mathcal{CN}\left( {0,k_1^2} \right)$ denotes the hardware distortion noise caused by the transmitter of $S_i$ and ${\tau _{rj}} \sim \mathcal{CN}\left( {0,k_2^2{P_i}|{h_{ij}}{|^2}} \right)$ denotes the hardware distortion noise caused by the receiver of $S_j$; ${{n_{ij}}}\sim \mathcal{CN}\left( {0,\sigma_{ij}^2} \right)$ is the additive white Gaussian noise (AWGN) at $S_j$. Hereby, the parameters $k_1$ and $k_2$ characterize the HIs levels of the transmitter and receiver respectively \cite{6630485}. Note that the product term in \eqref{3}, ${\sqrt {{P_i}} {h_{ij}}{\tau _{ti}}}$, follows a Gaussian distribution\footnote{Under the assumption of quasi-static fading channels, the channel fading coefficient $h_{ij}$ is constant in each transmission block \cite{6630485}. Meanwhile, the hardware distortion noise $\tau _{ti}$ caused by the transmitter of $S_i$ follows a Gaussian distribution with zero mean and variance $k^2_1$ \cite{8851300,7247469}. Thence, ${\sqrt {{P_i}} {h_{ij}}{\tau _{ti}}} \sim \mathcal{CN}\left( {0,k_1^2{P_i}|{h_{ij}}{|^2}} \right)$.} with zero mean and variance $k_1^2{P_i}|{h_{ij}}{|^2}$. For simplicity, we assume that $P_a=P_b=P_o$ and $\sigma _{ab}^2 = \sigma _{ba}^2 = {\sigma ^2}$ \cite{8613860}.
%

 Based on \eqref{3}, the SNDR of the direct link at $S_j$ is given as
 \begin{align}\label{4}
   {\gamma _{ij}} = \frac{{{\rho}|{h_{ij}}{|^2}}}{{\left( {k_1^2 + k_2^2} \right){\rho}|{h_{ij}}{|^2} + 1}},
 \end{align}
where $\rho  = \frac{{{P_o}}}{{{\sigma ^2}}}$ denotes the input SNR.

As shown in Fig. 3, the received signal at $R$ from $S_i$ ($i\!=\!a$ or $b$), under the presence of HIs, can be written as
\begin{align}\label{6}
  {y_{ir}} = {h_{ir}}\sqrt {{P_o}} \left( {{x_i} + {\tau _{ti}}} \right) + {n_{i{r_1}}},
\end{align}
where  ${n_{ir_1}} \sim \mathcal{CN}\left( {0,\sigma _{ir_1}^2} \right)$ denotes the antenna noise at $R$, ${\tau _{ti}} \sim \mathcal{CN}\left( {0,k_1^2} \right)$ denotes the hardware distortion noise caused by the transmitter of $S_i$.

Following the PS protocol, the received signal ${y_{ir}}$ is divided into two parts: $\sqrt \beta {y_{ir}}$ for EH and $\sqrt {1 - \beta } {y_{ir}}$ for IP. Thus, the total harvested energy
\footnote{For analytical tractability, this work considers the linear EH model \cite{8556567,7831382,7807356,7858707,7971948,7037438,8377371,8613860,8633928,8576644,7247469,7835665,8851300} instead of non-linear EH model \cite{7264986,7405266,8877296}. Please note that some non-linear EH models can be decomposed into piecewise linear EH models, where each segment of the EH model is linear and the results obtained based on linear EH models can be applied to each segment \cite{8647768,8913487}. Combining the derived results in this work and the Law of Total Probability, the system outage probability under the non-linear EH model can be obtained by following \cite{8647768,8913487}.
}
from both terminals can be expressed as \cite{7835665}
\begin{align}\label{7}
  {E_h} = \frac{T}{3}\eta \beta {P_o}\left( {|{h_{ar}}{|^2} + |{h_{br}}{|^2}} \right),
\end{align}
where $\eta\in(0,1)$ denotes the energy conversion efficiency at relay $R$. Here we ignore the energy harvested from the hardware distortion noise or the AWGN as it is much smaller than that from the RF signals transmitted by the two terminals \cite{8851300,7247469}.

Based on the \lq\lq harvest-then-forward\rq\rq\;scheme, the transmit power of $R$ is given as
\begin{align}\label{8}
  {P_r} = \frac{{{E_h}}}{{T/3}} = \eta \beta {P_o}\left( {|{h_{ar}}{|^2} + |{h_{br}}{|^2}} \right).
\end{align}
Furthermore, the received signal for IP is written as
\begin{align}\label{9}
{y_{\rm{R}}^{\rm{IP}}} = &{h_{ar}}\sqrt {\left( {1 - \beta } \right){P_o}} \left( {{x_a} + {\tau _{ta}}} \right)\nonumber\\
+& {h_{br}}\sqrt {\left( {1 - \beta } \right){P_o}} \left( {{x_b} + {\tau _{tb}}} \right)+ {\tau _{rr}} + {n_{ar}} + {n_{br}},
\end{align}
where ${\tau _{rr}} \sim \mathcal{CN} \left( {0,k_2^2(1-\beta){P_o}\left( {|{h_{ar}}{|^2} + |{h_{br}}{|^2}} \right)} \right)$ denotes the hardware distortion noise\footnote{In practice, the majority of the distortion noises at receiver is considered to be caused after the down-conversion process of the received antenna signal to baseband. Considering the complexity of theoretical analysis, the distortion noises cased by receiving process are neglected.} caused by the receiver of $R$, ${n_{ir}} \sim \mathcal{CN}\left( {0,\sigma _{ir}^2} \right)$ for $i \in \left\{ {a,b} \right\}$ denotes the AWGN at $R$.

In the RL phase, the relay firstly amplifies ${y_{\rm{R}}^{\rm{IP}}}$ using an amplification gain $G$  and broadcasts the amplified signal to both terminals. The received signal at the node  $S_i$ from the relay can be written as
 \begin{align}\label{10}
{y_{ri}} = {h_{ri}}Gy_{\rm{R}}^{{\rm{IP}}} + {h_{ri}}\sqrt {{P_r}} {\tau _{tr}} + {\tau _{ri}} + {n_{ri}},
\end{align}
 where $G \approx \sqrt {\frac{{{P_r}}}{{\left( {1 - \beta } \right)\left( {1 + k_1^2 + k_2^2} \right){P_o}\left( {|{h_{ar}}{|^2} + |{h_{br}}{|^2}} \right)}}}$ \cite{8851300}, ${\tau _{tr}} \sim \mathcal{CN}\left( {0,k_1^2} \right)$ denotes the distortion noise caused by the transmitter of $R$, ${\tau _{ri}} \sim \mathcal{CN}\left( {0,k_2^2{P_r}\left( {|{h_{ri}}{|^2}} \right)} \right)$ denotes the distortion noise caused by the receiver of $S_i$ and ${n_{ri}} \sim \mathcal{CN}\left( {0,\sigma _{ri}^2} \right)$ denotes the AWGN at $S_i$. For simplicity, we assume that $\sigma _{ar}^2 = \sigma _{br}^2 = \sigma _{ra}^2 = \sigma _{rb}^2 = {\sigma ^2}$ \cite{8613860}.

 After self-interference cancellation, the SNDR of the relaying link\footnote{Our derivations under the TDBC protocol can be applied to SWIPT based TWRNs with MABC, because under the MABC protocol, the SNDR of the relaying link  at $S_i$ can be written as ${\gamma _{ri}} = \frac{{{I_1}{\rho}|{h_{ir}}{|^2}|{h_{jr}}{|^2}}}{{{I_2}{\rho}|{h_{ir}}{|^2}\left( {|{h_{ar}}{|^2} + |{h_{br}}{|^2}} \right) + {I_3}|{h_{ir}}{|^2}/2 + 1}}$.} at $S_i$ can be calculated as
 \begin{align}\label{11}
 {\gamma _{ri}} = \frac{{{I_1}{\rho}|{h_{ir}}{|^2}|{h_{jr}}{|^2}}}{{{I_2}{\rho}|{h_{ir}}{|^2}\left( {|{h_{ar}}{|^2} + |{h_{br}}{|^2}} \right) + {I_3}|{h_{ir}}{|^2} + 1}},
 \end{align}
 with
 \begin{align}\notag
 {I_1} &= \frac{{\eta \beta }}{{1{\rm{ + }}k_1^2{\rm{ + }}k_2^2}},\\ \notag
 {I_2} &= \eta \beta \left( {\frac{{k_1^2 + k_2^2}}{{1{\rm{ + }}k_1^2{\rm{ + }}k_2^2}} + k_1^2 + k_2^2} \right),\\ \notag
 {I_3}& = \frac{{2\eta \beta }}{{\left( {1 - \beta } \right)\left( {1{\rm{ + }}k_1^2{\rm{ + }}k_2^2} \right)}}.
 \end{align}

\begin{figure*}[!t]
\normalsize
\setcounter{equation}{17}
\begin{align}\label{18}
{\mathbb{P}_{2}^1} =& 1 - \frac{2}{{\Gamma \left( {{m_a}} \right)\theta _a^{{m_a}}}}{e^{ - \frac{{{I_3}{I_4}{\gamma _{th}}}}{{{\theta _b}}}}}{\sum\limits_{l = 0}^{{m_b} - 1} {\sum\limits_{s = 0}^l {\sum\limits_{t = 0}^{l - s} {\left( \begin{array}{l}
l\\
s
\end{array} \right)\left( \begin{array}{l}
l - s\\
t
\end{array} \right)\frac{1}{{l!}}{{\left( {{I_2}\rho } \right)}^s}I_3^{l - s - t}{{\left( {\frac{{{I_4}{\gamma _{th}}}}{{{\theta _b}}}} \right)}^l}\left( {\frac{{{I_4}{\gamma _{th}}{\theta _a}}}{{{I_2}{I_4}\rho {\gamma _{th}}{\theta _a} + {\theta _b}}}} \right)} } } ^{\frac{{s + {m_a} - t}}{2}}}\nonumber\\
&\times {K_{s + {m_a} - t}}\left( {2\sqrt {\left( {\frac{{{I_4}{\gamma _{th}}}}{{{\theta _b}}}} \right)\left( {\frac{{{I_2}{I_4}\rho {\gamma _{th}}}}{{{\theta _b}}} + \frac{1}{{{\theta _a}}}} \right)} } \right).
\end{align}
\hrulefill
\setcounter{equation}{19}
\begin{align}\label{aa}
{\mathbb{P}_{3}^1} =& 1 - \frac{2}{{\Gamma \left( {{m_b}} \right)\theta _b^{{m_b}}}}{e^{ - \frac{{{I_3}{I_4}{\gamma _{th}}}}{{{\theta _a}}}}}{\sum\limits_{l = 0}^{{m_a} - 1} {\sum\limits_{s = 0}^l {\sum\limits_{t = 0}^{l - s} {\left( \begin{array}{l}
l\\
s
\end{array} \right)\left( \begin{array}{l}
l - s\\
t
\end{array} \right)\frac{1}{{l!}}{{\left( {{I_2}\rho } \right)}^s}I_3^{l - s - t}{{\left( {\frac{{{I_4}{\gamma _{th}}}}{{{\theta _a}}}} \right)}^l}\left( {\frac{{{I_4}{\gamma _{th}}{\theta _b}}}{{{I_2}{I_4}\rho {\gamma _{th}}{\theta _b} + {\theta _a}}}} \right)} } } ^{\frac{{s + {m_b} - t}}{2}}}\nonumber\\
&\times {K_{s + {m_b} - t}}\left( {2\sqrt {\left( {\frac{{{I_4}{\gamma _{th}}}}{{{\theta _a}}}} \right)\left( {\frac{{{I_2}{I_4}\rho {\gamma _{th}}}}{{{\theta _a}}} + \frac{1}{{{\theta _b}}}} \right)} } \right).
\end{align}
\setcounter{equation}{10}
\hrulefill
\end{figure*}

 Based on \eqref{4} and \eqref{11}, the end-to-end SNDRs after selection combination at $S_a$ and $S_b$ can be written respectively as
 \begin{align}\label{12}
   {\gamma _a} = \max \left\{ {{\gamma _{ba}},\;{\gamma _{ra}}} \right\},
 \end{align}
 \begin{align}\label{13}
   {\gamma _b} = \max \left\{ {{\gamma _{ab}},\;{\gamma _{rb}}} \right\}.
 \end{align}

 Note that ${\gamma _{ab}=\gamma _{ba}}$ due to the reciprocity of the $S_a$ to $S_b$ channel.

\section{System Outage Probability Analysis}
In this section, we derive a closed-form expression for the system outage probability and identify two crucial ceiling effects, viz., RCC and OSC. In addition, we analyze the diversity gain of our considered network.

\subsection{System Outage Probability}
The considered network will be in outage when either the transmission rate $\rm R_{a}$ from $S_a$ to $S_b$ or the transmission rate $\rm R_b$ from $S_b$ to $S_a$ is less than a given threshold $\rm R_{th}$. Thus, the system outage probability\footnote{From \eqref{7}, we can see that relay $R$ may harvest insufficient energy $E_b$ if the two terminals transmit at a low power level. According to (7)-(10), this  will lead to a low SNDR at both terminals and a high  system outage probability.}, $\mathbb{P}_{\rm out}$, can be expressed as
\begin{align}\label{14}
  {\mathbb{P}_{\rm out}} =& \Pr \left( {\min \left( \rm {{R_a},\;\rm {R_b}} \right) < {\rm R_{th}}} \right)\nonumber\\
   =& \Pr \left( {\min \left( {{\gamma _a},\;{\gamma _b}} \right) < {\gamma _{th}}} \right)\nonumber\\
   =& \Pr \left( {{\gamma _{ab}} < {\gamma _{th}}} \right)\big( \Pr \left( {{\gamma _{ra}} < {\gamma _{th}}} \right) + \Pr \left( {{\gamma _{rb}} < {\gamma _{th}}} \right)\nonumber\\
   &- \Pr \left( {{\gamma _{ra}} < {\gamma _{th}},{\gamma _{rb}} < {\gamma _{th}}} \right) \big).
\end{align}
where  ${\gamma _{th}} = {2^{{{3{{\rm{R}}_{{\rm{th}}}}} \mathord{\left/
 {\vphantom {{3{{\rm{R}}_{{\rm{th}}}}} T}} \right.
 \kern-\nulldelimiterspace} T}}} - 1$ denotes the SNDR threshold and ${\rm {R}}_i=\frac{T}{3}{\log _2}\left( {1 + {\gamma _i}} \right)$, $i \in \{ a,b\}$.
\begin{figure*}[!t]
\normalsize
\setcounter{equation}{21}
\begin{align}\label{22}
\mathbb{P}_{4}^1 =& 1 - \frac{2}{{\Gamma \left( {{m_a}} \right)\theta _a^{{m_a}}}}{e^{ - \frac{{{I_3}{I_4}{\gamma _{th}}}}{{{\theta _b}}}}}\sum\limits_{l = 0}^{{m_b} - 1} {\sum\limits_{s = 0}^l {\sum\limits_{t = 0}^{l - s} {\left( \begin{array}{l}
l\\
s
\end{array} \right)\left( \begin{array}{l}
l - s\\
t
\end{array} \right)\frac{1}{{l!}}{{\left( {{I_2}\rho } \right)}^s}I_3^{l - s - t}{{\left( {\frac{{{I_4}{\gamma _{th}}}}{{{\theta _b}}}} \right)}^l}{{\left( {\frac{{{I_4}{\gamma _{th}}{\theta _a}}}{{{I_2}{I_4}\rho {\gamma _{th}}{\theta _a} + {\theta _b}}}} \right)}^{^{\frac{{s + {m_a} - t}}{2}}}}} } }\nonumber\\
 &\times {K_{s + {m_a} - t}}\left( {2\sqrt {\left( {\frac{{{I_4}{\gamma _{th}}}}{{{\theta _b}}}} \right)\left( {\frac{{{I_2}{I_4}\rho {\gamma _{th}}}}{{{\theta _b}}} + \frac{1}{{{\theta _a}}}} \right)} } \right) - \frac{2}{{\Gamma \left( {{m_b}} \right)\theta _b^{{m_b}}}}{e^{ - \frac{{{I_3}{I_4}{\gamma _{th}}}}{{{\theta _a}}}}}\sum\limits_{l = 0}^{{m_a} - 1} {\sum\limits_{s = 0}^l {\sum\limits_{t = 0}^{l - s} {\left( \begin{array}{l}
l\\
s
\end{array} \right)\left( \begin{array}{l}
l - s\\
t
\end{array} \right)\frac{1}{{l!}}{{\left( {{I_2}\rho } \right)}^s}} } }\nonumber\\
&\times I_3^{l - s - t} {\left( {\frac{{{I_4}{\gamma _{th}}}}{{{\theta _a}}}} \right)^l}{\left( {\frac{{{I_4}{\gamma _{th}}{\theta _b}}}{{{I_2}{I_4}\rho {\gamma _{th}}{\theta _b} + {\theta _a}}}} \right)^{^{\frac{{s + {m_b} - t}}{2}}}}{K_{s + {m_b} - t}}\left( {2\sqrt {\left( {\frac{{{I_4}{\gamma _{th}}}}{{{\theta _a}}}} \right)\left( {\frac{{{I_2}{I_4}\rho {\gamma _{th}}}}{{{\theta _a}}} + \frac{1}{{{\theta _b}}}} \right)} } \right).
\end{align}
\hrulefill
\setcounter{equation}{24}
\hrulefill
\end{figure*}

Define $|{h_{ar}}{|^2} = X$, $|{h_{br}}{|^2} = Y$ and $|{h_{ab}}{|^2} = Z$.
Then  the system outage
probability $\mathbb{P}_{\rm out}$ can be re-expressed as
\setcounter{equation}{13}
\begin{align}
&{\mathbb{P}_{\rm out}} = \underbrace {\Pr \left( {\left( {1 - \left( {k_1^2 + k_2^2} \right){\gamma _{th}}} \right)\rho Z < {\gamma _{th}}} \right)}_{{\mathbb{P}_1}}\nonumber \\
 &\times \Bigg(\underbrace {\Pr \left( {\left( {{I_1} - {\gamma _{th}}{I_2}} \right)\rho Y < \rho {\gamma _{th}}{I_2}X + {I_3}{\gamma _{th}} + \frac{{{\gamma _{th}}}}{X}} \right)}_{{\mathbb{P}_2}}\nonumber\\
  & + \underbrace {\Pr \left( {\left( {{I_1} - {\gamma _{th}}{I_2}} \right)\rho X < \rho {\gamma _{th}}{I_2}Y + {I_3}{\gamma _{th}} + \frac{{{\gamma _{th}}}}{Y}} \right)}_{{\mathbb{P}_3}}\nonumber
\end{align}
\begin{align}\label{15}
 &- \underbrace {\Pr \left( \begin{array}{l}
\left( {{I_1} - {\gamma _{th}}{I_2}} \right)\rho Y < \rho {\gamma _{th}}{I_2}X + {I_3}{\gamma _{th}} + \frac{{{\gamma _{th}}}}{X},\\
\left( {{I_1} - {\gamma _{th}}{I_2}} \right)\rho X < \rho {\gamma _{th}}{I_2}Y + {I_3}{\gamma _{th}} + \frac{{{\gamma _{th}}}}{Y}
\end{array} \right)}_{\mathbb{P}_4}\Bigg).
\end{align}

Based on \eqref{2} and \eqref{25}, we derive the four probabilities on the right-hand side of \eqref{15} to obtain the closed-form expression for the system outage probability $\mathbb{P}_{\rm out}$, respectively. The first term, $\mathbb{P}_1$, can be calculated as
\begin{align}\label{16}
  {\mathbb{P}_1} = \left\{ \begin{array}{l}
{\mathbb{P}_{1}^1}\;,\;\;{\gamma _{th}} < \frac{1}{{k_1^2 + k_2^2}},\\
\;1\;\;,\;\;{\gamma _{th}} \ge \frac{1}{{k_1^2 + k_2^2}},
\end{array} \right.
\end{align}
where ${\mathbb{P}_{1}^1}$ is given by
\begin{align}\label{17}
  {\mathbb{P}_{1}^1} =& 1 - {e^{ - \frac{{{\gamma _{th}}}}{{{\theta _d}\rho \left( {1 - {\gamma _{th}}\left( {k_1^2 + k_2^2} \right)} \right)}}}}\nonumber\\
  &\times \sum\limits_{l = 0}^{{m_d}-1} {\frac{1}{{l!}}} {\left( {\frac{{{\gamma _{th}}}}{{{\theta _d}\rho \left( {1 - {\gamma _{th}}\left( {k_1^2 + k_2^2} \right)} \right)}}} \right)^l}.
\end{align}
\begin{proof}
See Appendix A.
\end{proof}
\begin{remark}
  In the considered TWRNs, $\mathbb{P}_1$ denotes the outage probability of the direct link between $S_a$ and $S_b$. As exhibited in \eqref{16}, due to the effects of HIs, the direct link will be in outage when $\gamma_{th}$ exceeds a certain value, i.e., ${\gamma _{th}} \ge \frac{1}{{k_1^2 + k_2^2}}$, where the maximum allowed value ${\gamma _{th}^{\max }} \approx \frac{1}{{k_1^2 + k_2^2}}$ decreases as the levels of HIs increase. This is intuitive since the instantaneous SNDR for the direct link in \eqref{4} is upper bounded by ${\gamma _{ij}} < \frac{1}{{k_1^2 + k_2^2}}$, $i,j \in \{ a,b\}$ and $i \ne j$. 
\end{remark}

The second term, $\mathbb{P}_2$ can be calculated as
\begin{align}\label{20}
{\mathbb{P}_2} = \left\{ {\begin{array}{*{20}{l}}
{{\mathbb{P}_{2}^1}\;,\;\;{\gamma _{th}} < \frac{{{I_1}}}{{{I_2}}}},\\
{\;1\;\;,\;\;{\gamma _{th}} \ge \frac{{{I_1}}}{{{I_2}}},}
\end{array}} \right.
\end{align}
where ${\mathbb{P}_{2}^1}$ is shown in \eqref{18} at the top of this page.

\begin{proof}
See Appendix B.
\end{proof}

Similarly, the third term, $\mathbb{P}_3$, can be calculated as
\begin{align}\label{21}
\setcounter{equation}{18}
  {\mathbb{P}_3} = \left\{ {\begin{array}{*{20}{l}}
{{\mathbb{P}_{3}^1}\;,\;\;{\gamma _{th}} < \frac{{{I_1}}}{{{I_2}}}},\\
{\;1\;\;,\;\;{\gamma _{th}} \ge \frac{{{I_1}}}{{{I_2}}},}
\end{array}} \right.
\end{align}
where ${\mathbb{P}_{3}^1}$ is given in \eqref{aa} at the top of this page.
\begin{remark}
  The terms $\mathbb{P}_2$ and $\mathbb{P}_3$ denote the T2T outage probabilities of the ${S_b}\mathop  \to \limits^R {S_a}$ and ${S_a}\mathop  \to \limits^R {S_b}$ links, respectively. As shown in \eqref{20} and \eqref{21}, the HIs pose undesirable constraint on $\gamma_{th}$, which inhibits information transmission from the relaying links when $\gamma_{th}$ exceeds a certain value, viz., ${\gamma _{th}} \ge \frac{{{I_1}}}{{{I_2}}} = \frac{1}{{\left( {k_1^2 + k_2^2} \right)\left( {2 + k_1^2 + k_2^2} \right)}}$. This is because the instantaneous SNDRs of the ${S_b}\mathop  \to \limits^R {S_a}$ and ${S_a}\mathop  \to \limits^R {S_b}$ links, given in \eqref{11}, exist an upper bound, i.e, ${\gamma _{ri}} < \frac{{{I_1}}}{{{I_2}}}$. Similarly, the maximum allowed ${\gamma _{th}^{\max }} \approx \frac{1}{{\left( {k_1^2 + k_2^2} \right)\left( {2 + k_1^2 + k_2^2} \right)}}$ decreases as the levels of HIs increase.
\end{remark}

The fourth term, $\mathbb{P}_4$, can be calculated as
\setcounter{equation}{20}
\begin{align}\label{56}
{\mathbb{P}_4} = \left\{ \begin{array}{l}
\mathbb{P}_4^ *\; ,\;\;{\gamma _{th}} < \frac{{{I_1}}}{{2{I_2}}},\\
\mathbb{P}_4^1\;,\;\;\frac{{{I_1}}}{{2{I_2}}} \le {\gamma _{th}} < \frac{{{I_1}}}{{{I_2}}},\\
\;1\;\;,\;\;{\gamma _{th}} \ge \frac{{{I_1}}}{{{I_2}}}.
\end{array} \right.
\end{align}
In \eqref{56}, $\mathbb{P}_4^1$ is shown in \eqref{22} at the top of the this page; $\mathbb{P}_4^ *$ equals either $\mathbb{P}^2_4$ or $\mathbb{P}^3_4$, as shown respectively in \eqref{29a} and \eqref{29b} at the top of the next page.
 $\mathbb{P}^2_4$ corresponds to the case that the quartic function (C.3) in Appendix C have only one real root, and $\mathbb{P}^3_4$ corresponds to the case that the quartic function (C.3) have three different positive real roots.
\begin{proof}
See Appendix C.
\end{proof}

\begin{figure*}[!t]
\normalsize
\setcounter{equation}{22}
\begin{align}\label{29a}
&{\mathbb{P}_4^2} = 1 - \frac{1}{{\Gamma \left( {{m_a}} \right)\theta _a^{{m_a}}}}\sum\limits_{l = 0}^{{m_b} - 1} {\frac{1}{{l!}}} {\left( {\frac{1}{{{\theta _b}}}} \right)^l}{\left( {\frac{1}{{{\theta _a}}} + \frac{1}{{{\theta _b}}}} \right)^{ - \left( {l + {m_a}} \right)}}\Gamma \left( {l + {m_a},{x_{in}}\left( {\frac{1}{{{\theta _a}}} + \frac{1}{{{\theta _b}}}} \right)} \right) - \frac{{\pi {x_{in}}}}{{2N\Gamma \left( {{m_a}} \right)\theta _a^{{m_a}}}}\nonumber\\
 &\;\;\;\;\;\;\;\;\times {e^{ - \frac{{{\gamma _{th}}{I_3}{I_4}}}{{{\theta _b}}}}}{\sum\limits_{l = 0}^{{m_b} - 1} {\sum\limits_{n = 1}^N {\frac{1}{{l!}}\left( {\frac{{{I_4}{\gamma _{th}}}}{{{\theta _b}}}} \right)} } ^l}\sqrt {1 - v_n^2} {\left( {k_n^1} \right)^{{m_a} - 1}}{\left( {\rho {I_2}k_n^1 + {I_3} + \frac{1}{{k_n^1}}} \right)^l}{e^{ - \left( {\frac{{\rho {\gamma _{th}}{I_2}{I_4}}}{{{\theta _b}}} + \frac{1}{{{\theta _a}}}} \right)k_n^1 - \frac{{{\gamma _{th}}{I_4}}}{{{\theta _b}k_n^1}}}}\nonumber\\
 &\;\;\;\;\;\;\;\;- \frac{1}{{\Gamma \left( {{m_b}} \right)\theta _b^{{m_b}}}}\sum\limits_{l = 0}^{{m_a} - 1} {\frac{1}{{l!}}} {\left( {\frac{1}{{{\theta _a}}}} \right)^l}{\left( {\frac{1}{{{\theta _a}}} + \frac{1}{{{\theta _b}}}} \right)^{ - \left( {l + {m_b}} \right)}}\Gamma \left( {l + {m_b},{x_{in}}\left( {\frac{1}{{{\theta _a}}} + \frac{1}{{{\theta _b}}}} \right)} \right) - \frac{{\pi {x_{in}}}}{{2N\Gamma \left( {{m_b}} \right)\theta _b^{{m_b}}}}\nonumber\\
 &\;\;\;\;\;\;\;\;\times {e^{ - \frac{{{\gamma _{th}}{I_3}{I_4}}}{{{\theta _a}}}}}{\sum\limits_{l = 0}^{{m_a} - 1} {\sum\limits_{n = 1}^N {\frac{1}{{l!}}\left( {\frac{{{I_4}{\gamma _{th}}}}{{{\theta _a}}}} \right)} } ^l}\sqrt {1 - v_n^2} {\left( {k_n^1} \right)^{{m_b} - 1}}{\left( {\rho {I_2}k_n^1 + {I_3} + \frac{1}{{k_n^1}}} \right)^l}{e^{ - \left( {\frac{{\rho {\gamma _{th}}{I_2}{I_4}}}{{{\theta _a}}} + \frac{1}{{{\theta _b}}}} \right)k_n^1 - \frac{{{\gamma _{th}}{I_4}}}{{{\theta _a}k_n^1}}}}.
 \end{align}
 \hrulefill
 \begin{align}\label{29b}
\mathbb{P}_4^3 &= U\left( {{K_1} - {K_2}} \right)\Bigg(1 + \frac{1}{{\Gamma \left( {{m_a}} \right)\theta _a^{{m_a}}}}\sum\limits_{l = 0}^{{m_b} - 1} {\frac{1}{{l!}}{{\left( {\frac{1}{{{\theta _b}}}} \right)}^l}} {\left( {\frac{{{\theta _a} + {\theta _b}}}{{{\theta _a}{\theta _b}}}} \right)^{ - \left( {l + {m_a}} \right)}}\Gamma \left( {l + {m_a},\frac{{{x_{in}}\left( {{\theta _a} + {\theta _b}} \right)}}{{{\theta _a}{\theta _b}}}} \right)\nonumber\\
& + \frac{1}{{\Gamma \left( {{m_b}} \right)\theta _b^{{m_b}}}}\sum\limits_{l = 0}^{{m_a} - 1} {\frac{1}{{l!}}{{\left( {\frac{1}{{{\theta _a}}}} \right)}^l}} {\left( {\frac{{{\theta _a} + {\theta _b}}}{{{\theta _a}{\theta _b}}}} \right)^{ - \left( {l + {m_b}} \right)}}\Gamma \left( {l + {m_b},\frac{{{x_{in}}\left( {{\theta _a} + {\theta _b}} \right)}}{{{\theta _a}{\theta _b}}}} \right)\nonumber\\
&- \frac{{\pi \left( {{\Phi _1} - {x_{in}}} \right)}}{{2N\Gamma \left( {{m_a}} \right)\theta _a^{{m_a}}}}{e^{ - \frac{{{\gamma _{th}}{I_3}{I_4}}}{{{\theta _b}}}}}\sum\limits_{l = 0}^{{m_b} - 1} {\sum\limits_{n = 1}^N {\frac{1}{{l!}}{{\left( {\frac{{{I_4}{\gamma _{th}}}}{{{\theta _b}}}} \right)}^l}} } \sqrt {1 - v_n^2} {\left( {k_n^2} \right)^{{m_a} - 1}}{\left( {\rho {I_2}k_n^2 + {I_3} + \frac{1}{{k_n^2}}} \right)^l}{e^{ - \left( {\frac{{\rho {\gamma _{th}}{I_2}{I_4}}}{{{\theta _b}}} + \frac{1}{{{\theta _a}}}} \right)k_n^2 - \frac{{{\gamma _{th}}{I_4}}}{{{\theta _b}k_n^2}}}}\nonumber\\
&- \frac{{\pi \left( {{\Phi _2} - {x_{in}}} \right)}}{{2N\Gamma \left( {{m_b}} \right)\theta _b^{{m_b}}}}{e^{ - \frac{{{\gamma _{th}}{I_3}{I_4}}}{{{\theta _a}}}}}\sum\limits_{l = 0}^{{m_a} - 1} {\sum\limits_{n = 1}^N {\frac{1}{{l!}}{{\left( {\frac{{{I_4}{\gamma _{th}}}}{{{\theta _a}}}} \right)}^l}} } \sqrt {1 - v_n^2} {\left( {k_n^2} \right)^{{m_b} - 1}}{\left( {\rho {I_2}k_n^2 + {I_3} + \frac{1}{{k_n^2}}} \right)^l}{e^{ - \left( {\frac{{\rho {\gamma _{th}}{I_2}{I_4}}}{{{\theta _a}}} + \frac{1}{{{\theta _b}}}} \right)k_n^2 - \frac{{{\gamma _{th}}{I_4}}}{{{\theta _a}k_n^2}}}}\nonumber\\
&\!-\! \frac{{{\pi ^2}}}{{4N\Gamma \left( {{m_a}} \right)\theta _a^{{m_a}}}}\sum\limits_{l = 0}^{{m_b}\! -\! 1} {\sum\limits_{n = 1}^N {\frac{1}{{l!}}{{\left( {\frac{1}{{{\theta _b}}}} \right)}^l}} } \sqrt {1\! - \!v_n^2} {\left( {\tan k_n^3 \!+\! {\Phi _1}} \right)^{{m_a}\! -\! 1}}{G^l}\left( {\tan k_n^3 \!+\! {\Phi _1}} \right){\sec ^2}\left( {k_n^3} \right){e^{ - \left( {\frac{{G\left( {\tan k_n^3 + {\Phi _1}} \right)}}{{{\theta _b}}} + \frac{{\tan k_n^3 + {\Phi _1}}}{{{\theta _a}}}} \right)}}\nonumber\\
 &\!- \!\frac{{{\pi ^2}}}{{4N\Gamma \left( {{m_b}} \right)\theta _b^{{m_b}}}}\sum\limits_{l = 0}^{{m_a}\! -\! 1} {\sum\limits_{n = 1}^N {\frac{1}{{l!}}{{\left( {\frac{1}{{{\theta _a}}}} \right)}^l}} } \sqrt {1 \!- \!v_n^2} {\left( {\tan k_n^3\! +\! {\Phi _2}} \right)^{{m_b}\! - \!1}}{G^l}\left( {\tan k_n^3 \!+ \!{\Phi _2}} \right){\sec ^2}\left( {k_n^3} \right){e^{ - \left( {\frac{{G\left( {\tan k_n^3+ {\Phi _2}} \right)}}{{{\theta _a}}} + \frac{{\tan k_n^3 + {\Phi _2}}}{{{\theta _b}}}} \right)}}\Bigg)\nonumber\\
 & + (1 - U\left( {{K_1} - {K_2}} \right))\mathbb{P}_4^2.
\end{align}
\setcounter{equation}{24}
\hrulefill
\end{figure*}

Substituting \eqref{16}, \eqref{20}, \eqref{21} and \eqref{56} into \eqref{15}, the system outage probability, $\mathbb{P}_{\rm out}$, can be expressed as
\setcounter{equation}{24}
\begin{align}\label{32}
{\mathbb{P}_{\rm out}} = \left\{ \begin{array}{l}
\mathbb{P}_1^1\left( {\mathbb{P}_2^1 + \mathbb{P}_3^1 - \mathbb{P}_4^ * } \right),\;{\gamma _{th}} < \frac{{{I_1}}}{{2{I_2}}},\\
\;\;\;\;\;\;\;\;\;\mathbb{P}_1^1\;\;\;\;\;\;\;\;\;\;\;\;\;\;\;\;,\;\;\frac{{{I_1}}}{{2{I_2}}} \le {\gamma _{th}} < \frac{1}{{k_1^2 + k_2^2}},\\
\;\;\;\;\;\;\;\;\;\;1\;\;\;\;\;\;\;\;\;\;\;\;\;\;\;\;\;,\;\;{\gamma _{th}} \ge \frac{1}{{k_1^2 + k_2^2}},
\end{array} \right.
\end{align}
where the inequality $\frac{1}{{k_1^2 + k_2^2}} \ge \frac{{{I_1}}}{{{I_2}}} \ge \frac{{{I_1}}}{{2{I_2}}}$ always holds.
\begin{remark}
 The derived expressions in \eqref{32} can serve for the following purposes. Firstly, the closed-form expressions can be used to quantify the impact of HIs on the system outage probability instead of the Monte Carlo simulations. Secondly, based on \eqref{32}, we can determine two crucial ceiling effects in high data rate regions, which provide guidelines for practical system design. This will be discussed in the next subsection. Lastly, we can obtain some key insights into the effects of various system parameters on the system outage performance by numerically calculating \eqref{32}. Note that such an approach has been widely adopted in existing works \cite{8556567,7831382,7807356,7858707,7971948,7037438,8377371,8613860,8633928,8576644}.
\end{remark}

\subsection{Ceiling Effects}
 The first line on the right-hand side of \eqref{32}, $\mathbb{P}_1^1\left( {\mathbb{P}_2^1 + \mathbb{P}_3^1 - \mathbb{P}_4^ * } \right)$, accounts for the cooperative information exchange via both the direct and relaying links, whereas the second line, $\mathbb{P}_1^1$, arises from the information exchange via the direct link only. From \eqref{32}, we can observe that the HIs pose undesirable constraints on $\gamma_{th}$, which in turn causes ceiling effects on the considered network. Specifically, when ${\gamma _{th}} \ge \frac{{{I_1}}}{{2{I_2}}} = \frac{1}{{2\left( {k_1^2 + k_2^2} \right)\left( {2 + k_1^2 + k_2^2} \right)}}$, the relaying link ceases cooperative information exchange, and hence the outage performance of the system depends only on the direct link. This effect is referred to as RCC. When ${\gamma _{th}} \ge \frac{1}{{k_1^2 + k_2^2}}$, the information exchange through the direct link also fails and the system goes in outage, which is called OSC. Obviously, the thresholds of $\gamma_{th}$ corresponding to RCC and OSC are both determined by HIs levels. Note that RCC always appears before OSC due to $\frac{{{I_1}}}{{2{I_2}}} \le \frac{1}{{k_1^2 + k_2^2}}$. This indicates that the relaying link is more sensitive to transceiver HIs than the direct link.


\subsection{Diversity Gain}
The diversity gain of our considered network can be calculated as
\begin{align}
  d =  - \mathop {\lim }\limits_{\rho  \to \infty } \frac{{\log \left( {{\mathbb{P}_{\rm out}}} \right)}}{{\log \left( \rho  \right)}}.
\end{align}

Based on \eqref{32}, the diversity gain can be calculated separately for the following three cases.
\subsubsection{} When ${\gamma _{th}} \ge \frac{1}{{k_1^2 + k_2^2}}$, $d=0$.
\subsubsection{} When $\frac{{{I_1}}}{{2{I_2}}} \le {\gamma _{th}} < \frac{1}{{k_1^2 + k_2^2}}$, we have
\begin{align}\label{49}
  d &=  - \mathop {\lim }\limits_{\rho  \to \infty } \frac{{\log \left( {1 - {e^{ - \frac{a}{\rho }}} \sum\limits_{l = 0}^{{m_d} - 1} {\frac{1}{{l!}}} {{\left( {\frac{a}{\rho }} \right)}^l}} \right)}}{{\log \left( \rho  \right)}}\nonumber \\
  &\mathop  = \limits^{x = \frac{1}{\rho }} \mathop {\lim }\limits_{x \to 0} \frac{{\log \left( {1 - {e^{ - ax}} \sum\limits_{l = 0}^{{m_d} - 1} {\frac{1}{{l!}}} {{\left( {ax} \right)}^l}} \right)}}{{\log \left( x \right)}}\nonumber \\
  &= \mathop {\lim }\limits_{x \to 0} \frac{{\frac{1}{{\left( {{m_d} - 1} \right)!}}{{\left( {ax} \right)}^{{m_d}}}}}{{{e^{ax}} - \sum\limits_{l = 0}^{{m_d} - 1} {\frac{1}{{l!}}{{\left( {ax} \right)}^l}} }},
\end{align}
where $a = \frac{{{\gamma _{th}}}}{{{\theta _d}\left( {1 - {\gamma _{th}}\left( {k_1^2 + k_2^2} \right)} \right)}}$.

Using the Taylor series for ${{e^{ax}}}$, \eqref{49} can be re-expressed as
\begin{align}
  d &= \mathop {\lim }\limits_{x \to 0} \frac{{\frac{1}{{\left( {{m_d} - 1} \right)!}}{{\left( {ax} \right)}^{{m_d}}}}}{{\sum\limits_{l = 0}^\infty  {\frac{1}{{l!}}{{\left( {ax} \right)}^l}}  - \sum\limits_{l = 0}^{{m_d} - 1} {\frac{1}{{l!}}{{\left( {ax} \right)}^l}} }}\nonumber \\
  &= \mathop {\lim }\limits_{x \to 0} \frac{{\frac{1}{{\left( {{m_d} - 1} \right)!}}{{\left( {ax} \right)}^{{m_d}}}}}{{\sum\limits_{l = {m_d}}^\infty  {\frac{1}{{l!}}{{\left( {ax} \right)}^l}} }} = {m_d}.
\end{align}
\subsubsection{} When ${\gamma _{th}} < \frac{{{I_1}}}{{2{I_2}}}$, we have
\begin{align}\label{50}
  d &=  - \mathop {\lim }\limits_{\rho  \to \infty } \frac{{\log \left( \mathbb{P}_1^1\left( {\mathbb{P}_2^1 + \mathbb{P}_3^1 - \mathbb{P}_4^ * } \right) \right)}}{{\log \left( \rho  \right)}},\nonumber\\
  &= - \mathop {\lim }\limits_{\rho  \to \infty } \frac{{\log \left( \mathbb{P}_1^1 \right)}}{{\log \left( \rho  \right)}}  - \mathop {\lim }\limits_{\rho  \to \infty } \frac{{\log \left( {\mathbb{P}_2^1 + \mathbb{P}_3^1 - \mathbb{P}_4^ * } \right)}}{{\log \left( \rho  \right)}}\nonumber\\
  &= {m_d} - \mathop {\lim }\limits_{\rho  \to \infty } \frac{{\log \left( {\mathbb{P}_2^1 + \mathbb{P}_3^1 - \mathbb{P}_4^ * } \right)}}{{\log \left( \rho  \right)}}.
\end{align}
In what follows, we will prove $\mathop {\lim }\limits_{\rho  \to \infty } \frac{{\log \left( {\mathbb{P}_2^1 + \mathbb{P}_3^1 - \mathbb{P}_4^ * } \right)}}{{\log \left( \rho  \right)}}=0$.
Based on \eqref{51} in Appendix B, we have $\mathop {\lim }\limits_{\rho  \to \infty } {\mathbb{P}_2^1} = \int_0^\infty  {\int_0^{\frac{{{\gamma _{th}}{I_2}}}{{{I_1} - {\gamma _{th}}{I_2}}}x} {{f_X}\left( x \right){f_Y}\left( y \right)} } dydx = {C_1}$, where $0 < {C_1} < 1$. Similarly, $\mathop {\lim }\limits_{\rho  \to \infty } \mathbb{P}_3^1 = \int_0^\infty  {\int_0^{\frac{{{\gamma _{th}}{I_2}}}{{{I_1} - {\gamma _{th}}{I_2}}}y} {{f_Y}\left( y \right){f_X}\left( x \right)} } dxdy = {C_2}$, where $0 < {C_2} < 1$.
In addition, due to $\mathop {\lim }\limits_{\rho  \to \infty } {x_{in}} = 0$, the integral for $\mathbb{P}_4^*$ approaches 0 when ${\rho  \to \infty }$.

According to the above analysis, $d$ can be summarized as
\begin{align}\label{131}
  d = \left\{ \begin{array}{l}
{m_d}\;,\;{\gamma _{th}} < \frac{1}{{k_1^2 + k_2^2}},\\
\;0\;\;\;,\;{\gamma _{th}} \ge \frac{1}{{k_1^2 + k_2^2}}.
\end{array} \right.
\end{align}
%

\section{Simulation Results}
In this section, we present simulation results to validate the above theoretical analysis and evaluate the impacts of various parameters on the considered network.
Unless specifically stated, the simulation parameters are set as follows \cite{8633928,8364583,8851300}. We assume that $\eta  = 0.6$, $\beta=0.8$, $T = 1$ \!sec and ${k_1} = {k_2} = {k_{ave}}$.
The distances of $S_a-R$, $S_b-R$ and $S_a-S_b$ links are set as $d_{ar}=5$ m, $d_{br}=5$ m and $d_{ab}=10$ m, respectively. The average powers $\Omega_a$, $\Omega_b$ and $\Omega_{ab}$ are expressed as ${\Omega _a} = d_{ar}^{ - {\alpha _1}}$, ${\Omega _b} = d_{br}^{ - {\alpha _1}}$ and ${\Omega _d} = d_{ab}^{ - {\alpha _2}}$, where $\alpha_1=2.7$ and $\alpha_2=3$ denote the path loss exponents of the relaying link and the direct link, respectively.

Fig. 4(a) depicts the system outage probability versus the input SNR to verify the correctness of the derivations in all three cases of \eqref{32}.
Four different data rates are considered, viz., $1$~bit/Hz, $1.5$~bit/Hz, $1.75$~bit/Hz and $2$~bit/Hz, corresponding to  four SNDR thresholds, viz. $7$ dB, $22$ dB, $37$ dB and $63$ dB, respectively.
We also determine three threshold constraints caused by the HIs, viz., $\frac{{{I_1}}}{{2{I_2}}}=12$ dB, $\frac{{{I_1}}}{{{I_2}}}=24$ dB and $\frac{1}{{k_1^2 + k_2^2}} = 50$ dB.
Based on \eqref{32} and the used parameters, the system outage probability is computed as
\begin{align}\notag
{\mathbb{P}_{{\rm{out}}}} = \left\{ {\begin{array}{*{20}{c}}
{\mathbb{P}_1^1\left( {\mathbb{P}_2^1 + \mathbb{P}_3^1 -\mathbb{P}_4^*} \right),\;{\gamma _{th}} = 7\;{\rm{dB}}}\\
{\;\;\mathbb{P}_1^1,\;\;\;\;\;\;\;\;\;\;\;\;{\gamma _{th}} = 22\;{\rm{dB}}\;{\rm{or}}\;37\;{\rm{dB}}}\\
{1,\;\;\;\;\;\;\;\;\;\;\;\;\;\;\;\;\;\;\;\;\;\;\;\;\;{\gamma _{th}} = 63\;{\rm{dB}}}
\end{array}} \right..
\end{align}
One observation from Fig. 4(a) is that the analytical results (\lq $\triangledown$, +, $\Box$, $\lozenge$\rq \;marked results) match the simulation results (the solid line curves) well. This verifies the accuracy of the derived closed-form expression in \eqref{32}. Fig. 4(b) shows the relative approximate error against parameter $N$ (which determines the tradeoff between complexity and accuracy for the Gaussian-Chebyshev quadrature, as defined in \eqref{30} in Appendix C) to illustrate the accuracy of the Gaussian-Chebyshev quadrature approximation approach, which is used to obtain the approximate system outage probability in \eqref{32} when ${\gamma _{th}} < \frac{{{I_1}}}{{2{I_2}}}$.
\begin{figure}[!t]
\centering
\subfigure[]{
\includegraphics[width=5.5cm]{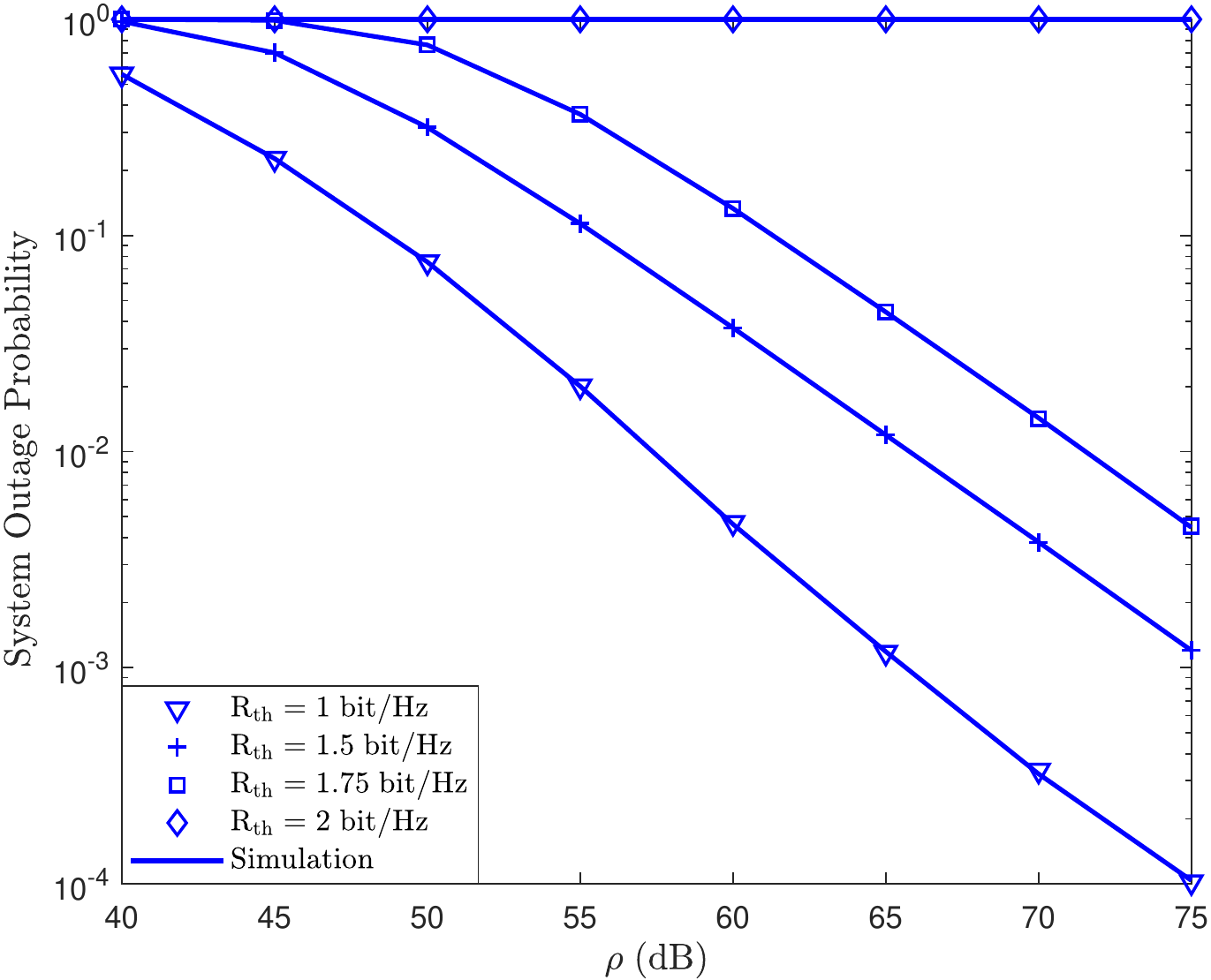}
}
\subfigure[]{
\includegraphics[width=5.5cm]{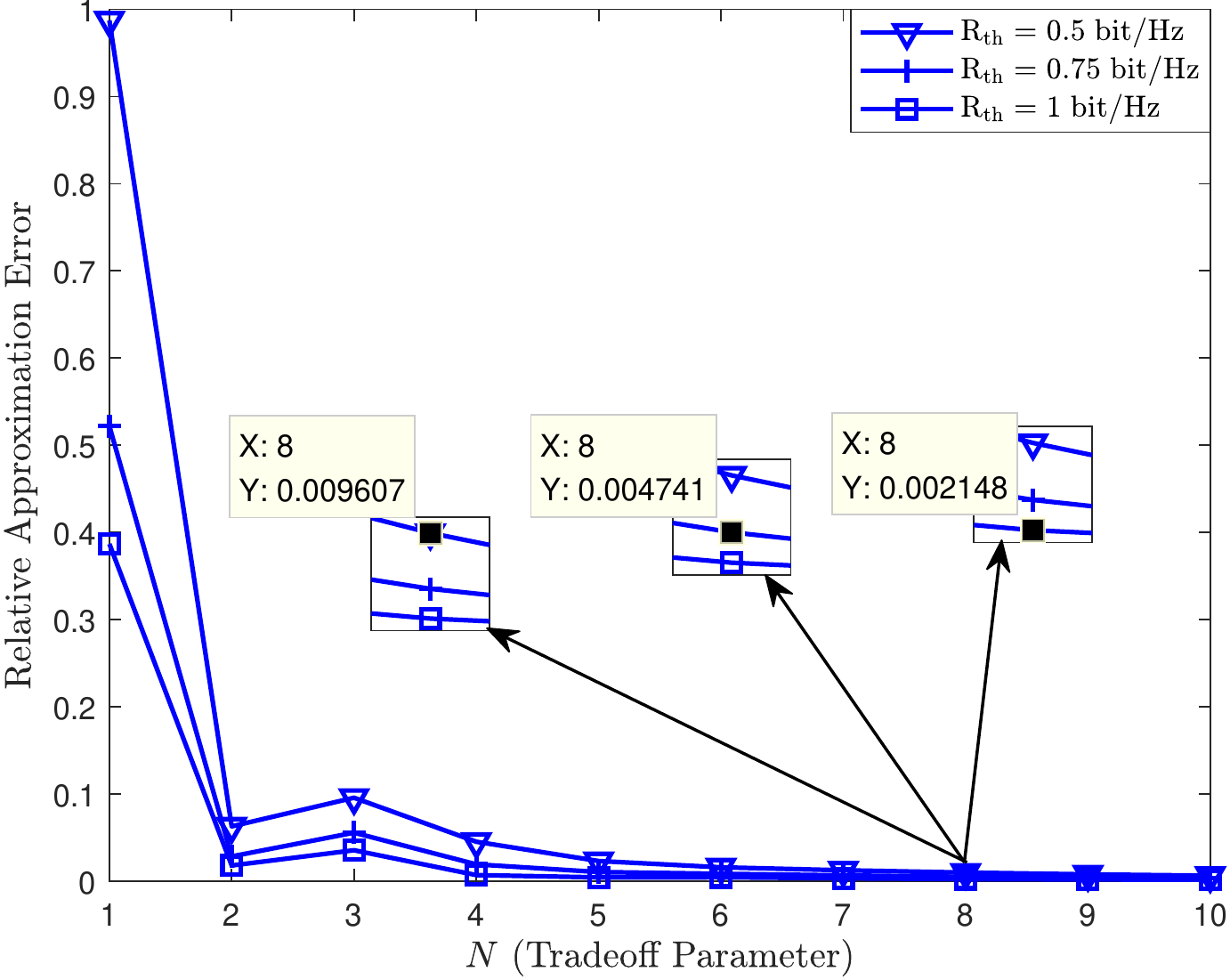}
}
\caption{Verification of the derived system outage probability, where $k_{ave}=0.1$ and $\{m_a, m_b, m_d\}=\{2, 2, 1\}$.}
\end{figure}
Accordingly, in Fig. 4(a), the Gaussian-Chebyshev quadrature is only used for the data rate of $1$~bit/Hz. In Fig. 4(b), we consider three data rates, viz., $0.5$~bit/Hz, $0.75$~bit/Hz and $1$~bit/Hz. Following \cite{8633928}, the relative approximate error is expressed as
\begin{align}\notag
\delta  = \left| {\frac{{{\rm{analytical\;result  -  simulation\;result}}}}{{{\rm{simulation\;result}}}}} \right|,
\end{align}
where the analytical and simulation results are obtained from \eqref{32} and Monte-Carlo simulations, respectively. As shown in Fig. 4(b), with the increase of $N$, $\delta$ approaches zero. For $N = 8$ and $\rm R_{th}=1$ bit/Hz, the corresponding $\delta$ is $0.002148$, which indicates a sufficient accuracy for the system outage probability. Therefore, the derived expression based on the Gaussian-Chebyshev approach in \eqref{32} can be used to evaluate the system outage performance of the considered network.

\begin{figure}[!t]
  \centering
  \includegraphics[width=0.45\textwidth]{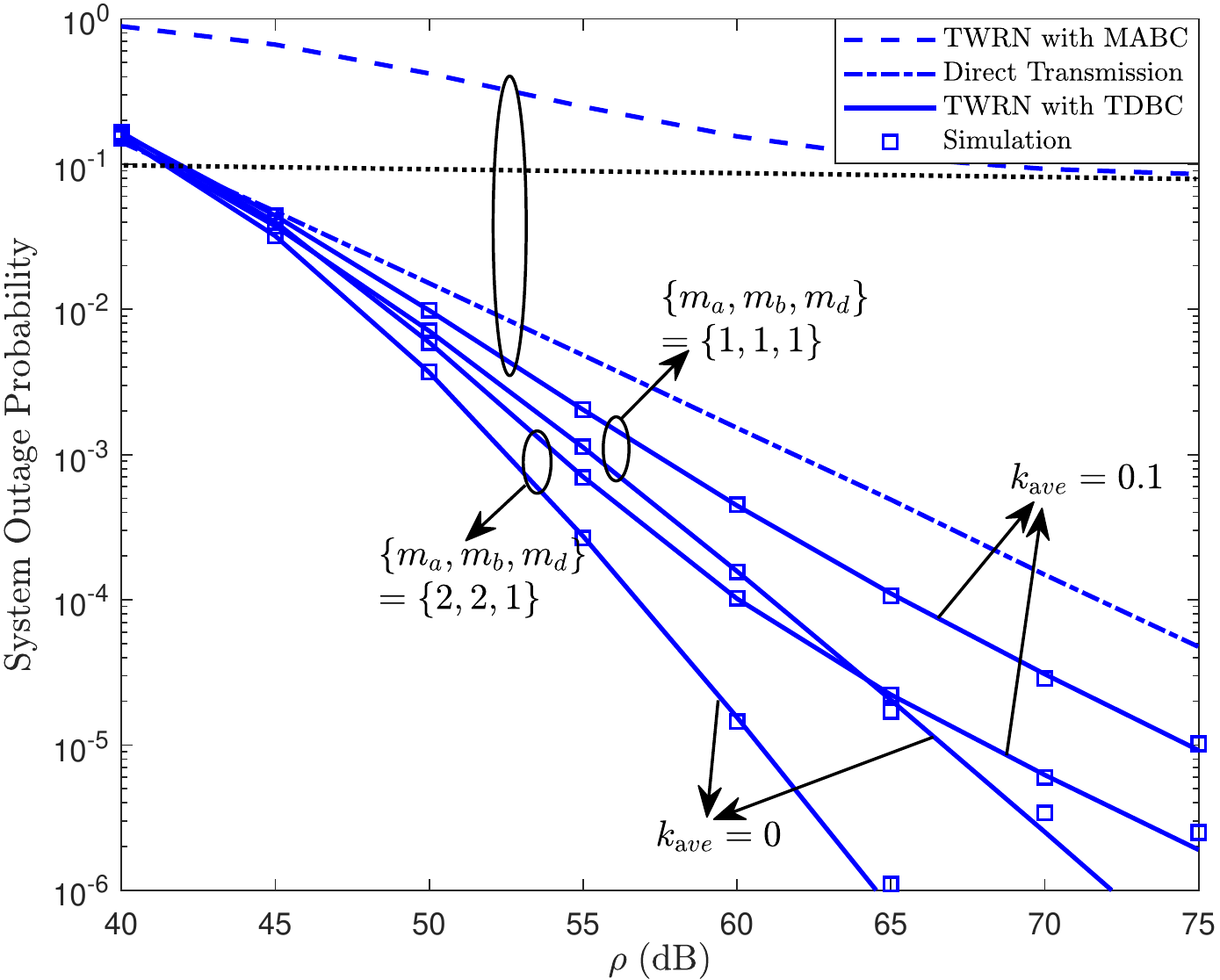}\\
  \caption{$\mathbb{P}_{\rm{out}}$ vs. $\rho$ for various HI levels and shape parameters, where $\rm R_{th}=0.5$ bit/Hz.}
\end{figure}

\begin{figure}[!t]
  \centering
  \includegraphics[width=0.45\textwidth]{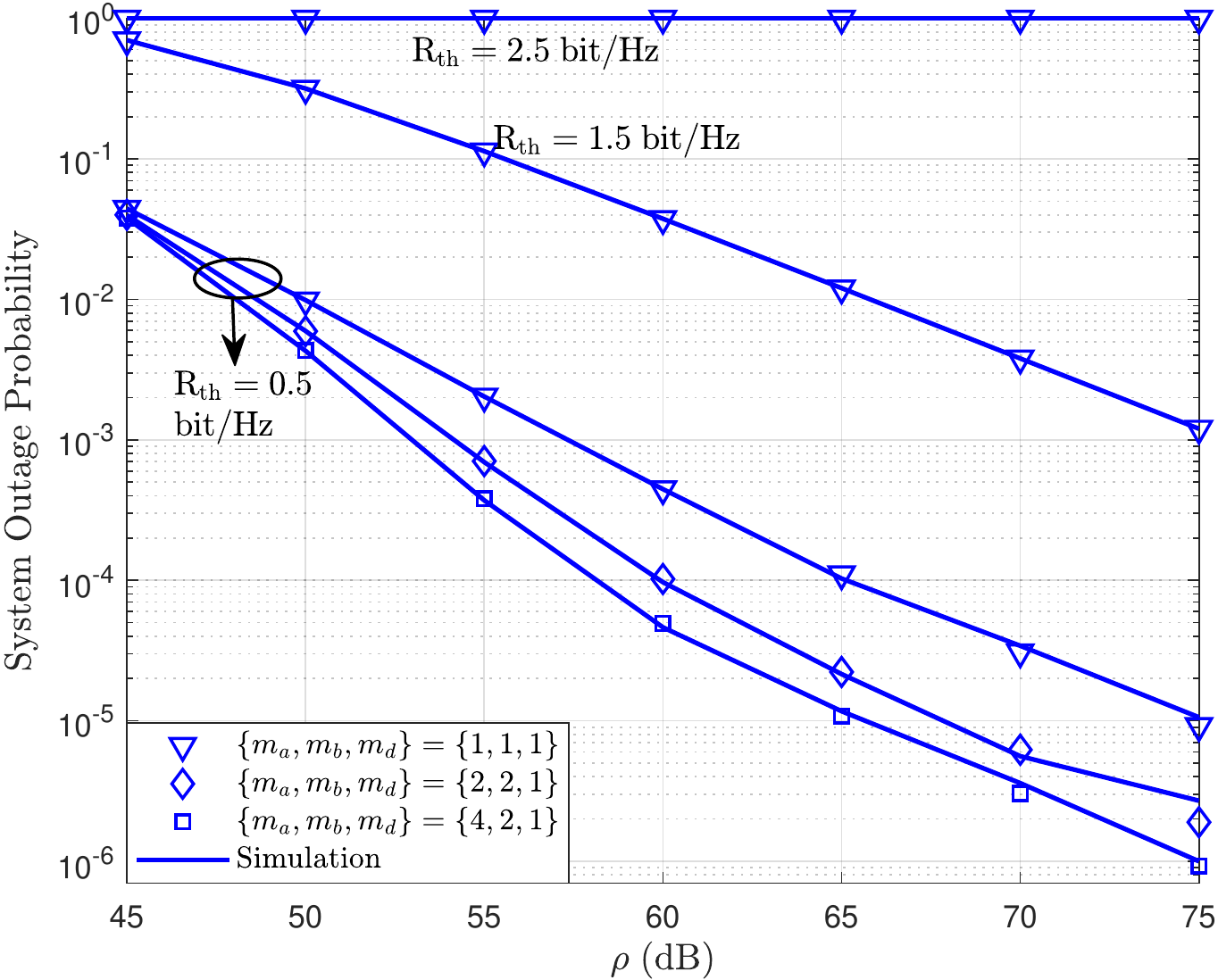}\\
  \caption{$\mathbb{P}_{\rm{out}}$ vs. $\rho$ for various data rate requirements, where $k_{ave}=0.1$.}
\end{figure}
Fig. 5 compares the system outage performance among three transmission protocols, namely the TDBC, the MABC, and the direct transmission, under different HIs levels and shape parameters of relaying links.
In the direct transmission protocol, each transmission block is divided into two equal phases. In the first phase, terminal $S_a$ transmits its signal to terminal $S_b$, while terminal $S_b$ transmit its signal to terminal $S_a$ in the second phase.
We set $\beta=0.8$ and $\rm R_{th}=0.5$ bit/Hz, where the corresponding SNDR threshold is maintained below that of OSC, i.e., $\gamma_{th}<\frac {1}{k^2_1+k^2_2}$.
For a fair comparison, the three transmission protocols assume the same energy consumption and the same transmission rate threshold at each terminal.
We can see that the TDBC achieves the lowest system outage probability while the system outage probability of the MABC is the highest.
This is because TDBC leverages both the direct and relaying links, but, the MABC and direct transmission rely only on the relaying link and the direct link, respectively.
For the given HIs levels, the system outage probability of the TWRN with TDBC decreases with the increase of shape parameters $m_a$ and $m_b$. This is because the relaying link quality improves with the increase of $m_a$ and $m_b$, leading to a reduced system outage probability. For the given shape parameters, the system outage probability of the TWRN with TDBC under HIs ($k_{ave}=0.1$) is much higher than that with ideal hardware ($k_{ave}=0$). This indicates that the HIs seriously deteriorates the system outage performance.

\begin{figure}[!t]
  \centering
  \includegraphics[width=0.44\textwidth]{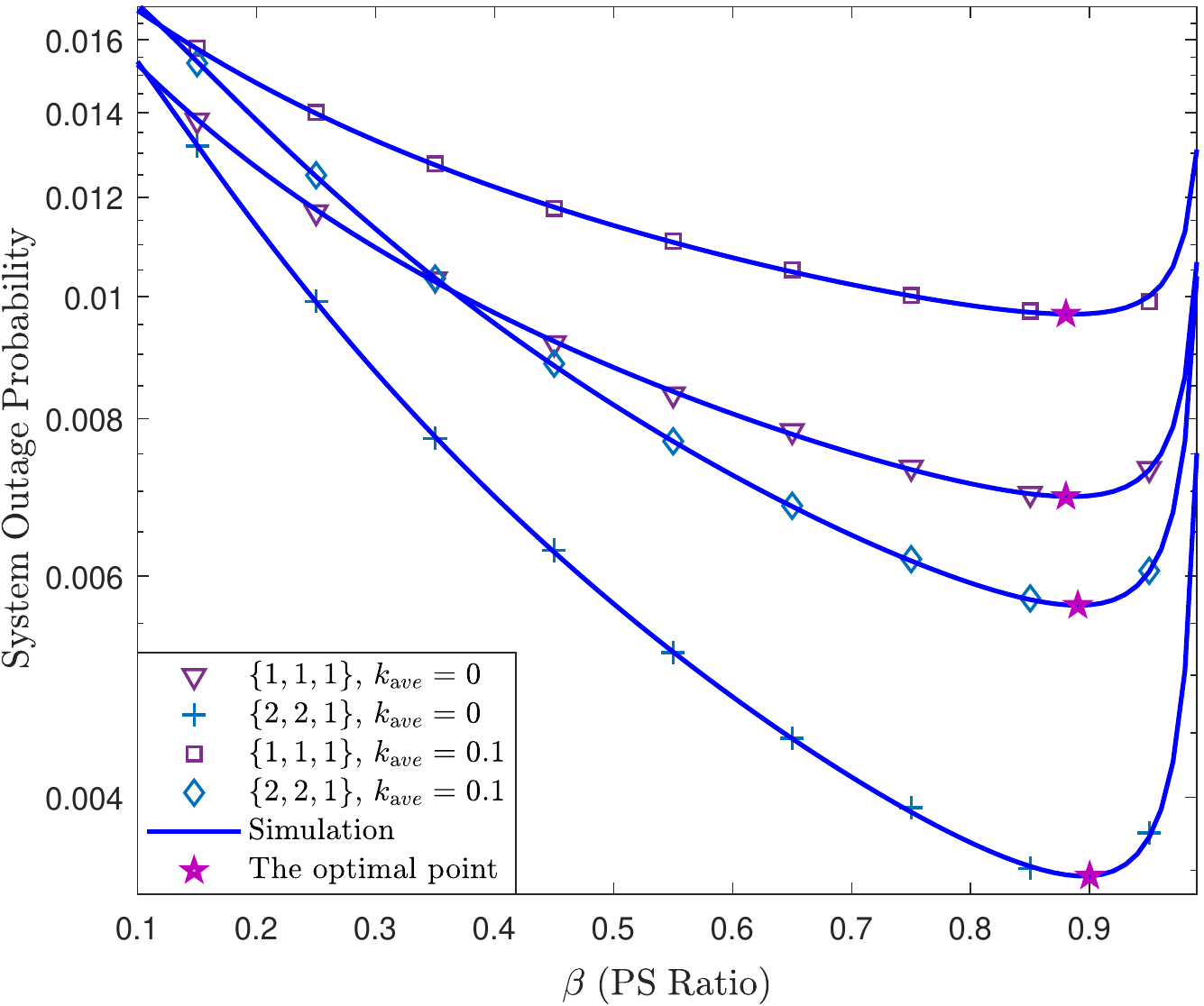}\\
  \caption{$\mathbb{P}_{\rm{out}}$ vs. $\beta$ for various HI levels and shape parameters, where $\rm R_{th}=0.5$ bit/Hz and $\rho=50$ dB.}
\end{figure}
\begin{figure}[!t]
  \centering
  \includegraphics[width=0.45\textwidth]{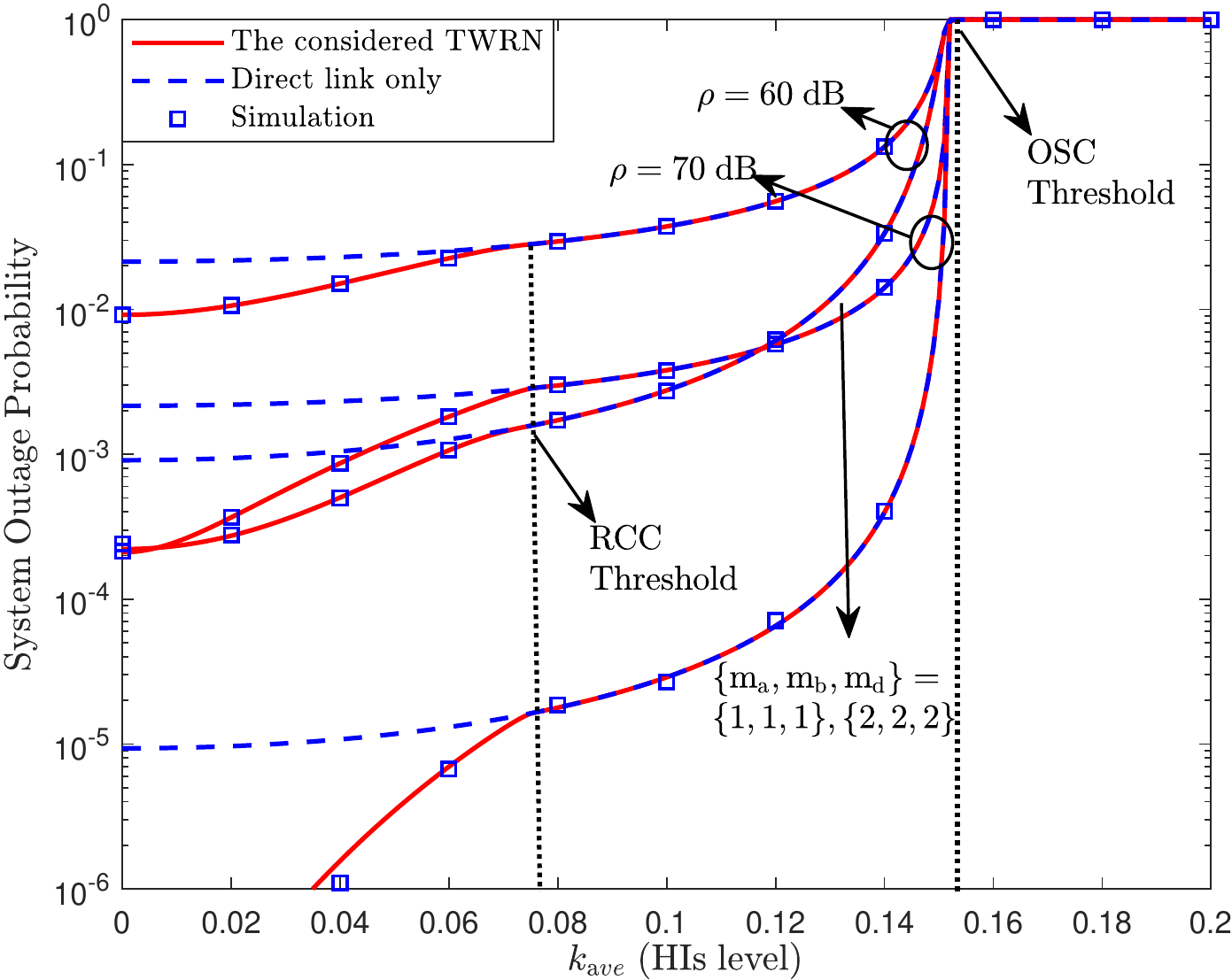}\\
  \caption{$\mathbb{P}_{\rm{out}}$ vs. $k_{ave}$ for various shape parameters, where $\rm R_{th}=1.5$ bps/Hz.}
\end{figure}
Fig. 6 validates the analysis of the diversity gain in Section III-C. We set the data rates as 0.5 bit/Hz, 1.5 bit/Hz and 2.5 bit/Hz, corresponding to third SNDR thresholds, i.e., 1.8 dB, 21.6 dB and 180 dB respectively. Also, $k_{ave}$ is set as 0.1, in this case, the thresholds for RCC and OSC are $\frac{{{I_1}}}{{2{I_2}}}=12$ dB and $\frac{1}{{k_1^2 + k_2^2}} = 50$ dB, respectively. Clearly, when ${\rm R_{th}}=2.5$ bit/Hz, the system outage probability equals to one and the diversity gain is zero, as expected in \eqref{131}. While for ${\rm R_{th}}=0.5$ or $1.5$ bit/s, the achievable diversity gain equals to $m_d$. In other words, the relaying link has no contribution to the diversity gain when the HIs exist.

Fig. 7 shows  the system outage probability with the increase of PS ratio $\beta$.
Firstly, as shown in this figure, the system outage probability  firstly decreases, and then  increases. Thus there is an optimal PS ratio to  minimize  the system outage probability. This phenomenon can be explained as following. When $\beta$ is smaller than the optimal PS ratio, the relay  harvests less power, leading to a small transmit power and a large system outage probability. When $\beta$ is greater than the optimal PS ratio, the more received signal power is wasted on EH and less power is used for the ${S_i} \to R$ link. Accordingly, a poor
signal strength and the harmful signal are amplified simultaneously, leading to a low SDNR at both terminals.
In addition, we can  observe that, for the given $k_{ave}$, the  shape parameters has little effect on optimal PS ratio. Particular, the optimal PS ratio increases with the shape parameters of relaying links.

Fig. 8 validates the RCC and OSC with different HIs levels, $k_{ave} \in [0,0.2]$. We set $\rm R_{th}=1.5$ bps/Hz, viz., $\gamma_{th}\approx22$ dB.
Based on the discussion in Section III-B and this figure,  we can  identify three hardware impairment regions as follows:

1) When $k_{ave} \in [0, 0.0758)$, the considered network can achieve better system outage performance duo to cooperation information exchange via both the direct and relaying links. Specially, $k_{ave}=0.0758$ is the RCC threshold with respect to HIs levels;

2) When $k_{ave} \in [0.0758, 0.152)$, although the system outage performance degrade because of the outage of relaying link, the information exchange can be achieved via direct link only. This suggests that the relaying link is more sensitive to HIs than the direct link. Specially, $k_{ave}=0.152$ is the OSC threshold;

3) When $k_{ave} \in [0.152, 0.2)$, no matter what value SNR takes, the system outage probability is always 1.

\begin{figure}[!t]
  \centering
  \includegraphics[width=0.45\textwidth]{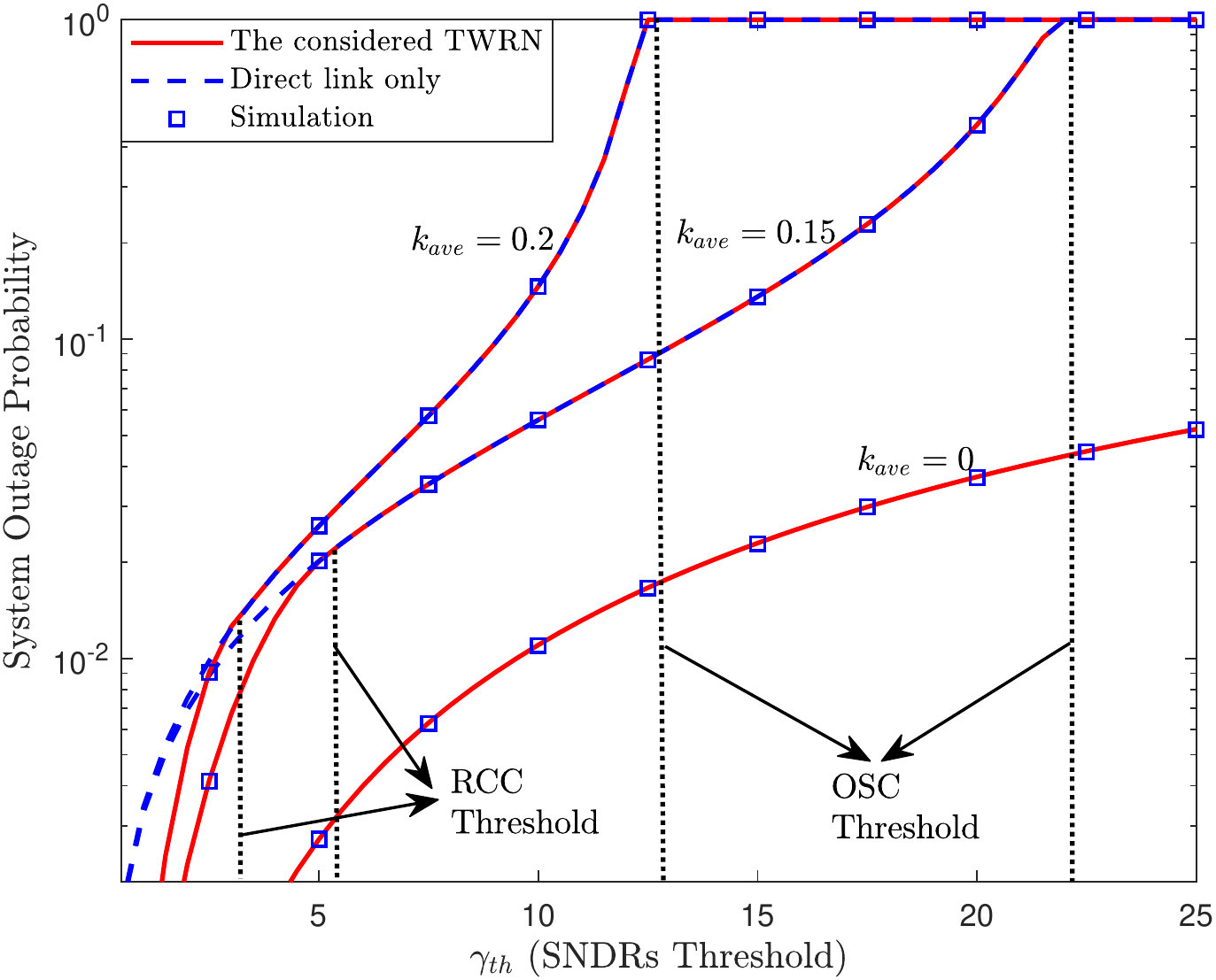}\\
  \caption{$\mathbb{P}_{\rm{out}}$ vs. $\gamma_{th}$ for various HIs levels, where $\rho=50$ dB.}\vspace{-12pt}
\end{figure}

Fig. 9 studies the impacts of $\gamma_{th}$ on the RCC and OSC with different HIs levels.
Let's take $k_{ave}=0.15$ as an example, for this, the RCC and OSC effects occur at the SNDRs threshold $\gamma_{th} \approx 5.4$ dB (${\gamma _{th}} \ge \frac{1}{{2\left( {k_1^2 + k_2^2} \right)\left( {2 + k_1^2 + k_2^2} \right)}}$) and $\gamma_{th} \approx 22.2$ dB (${\gamma _{th}} \ge \frac{1}{{k_1^2 + k_2^2}}$), respectively. Referring to the analysis in section III-B and Fig. 9, we can be sure that as $\gamma_{th}$ across the RCC threshold ($\gamma_{th}\geq 5.4$ dB), the system outage performance depends only on that of the direct link. Apparently, the direct link partially compensates for RCC effect ( in the threshold region $5.4$ dB$\leq\gamma_{th}<22.2$ dB) until the appearance of OSC effect. When $\gamma_{th}$ further exceeds the OSC threshold ($\gamma_{th}\geq 22.2$ dB), the system outage probability is equal to 1. Moreover, as shown in figure, the gap between the ideal hardware case ($k_{ave}=0$) and impairment cases ($k_{ave}=0.15$ and $0.2$.) is relatively small at low $\gamma_{th}$ region but other can be very large. Therefore, we can conclude that HIs are detrimental for system to keep with its outage performance standards, specially in high-data-rate applications.
\begin{figure}[!t]
  \centering
  \includegraphics[width=0.45\textwidth]{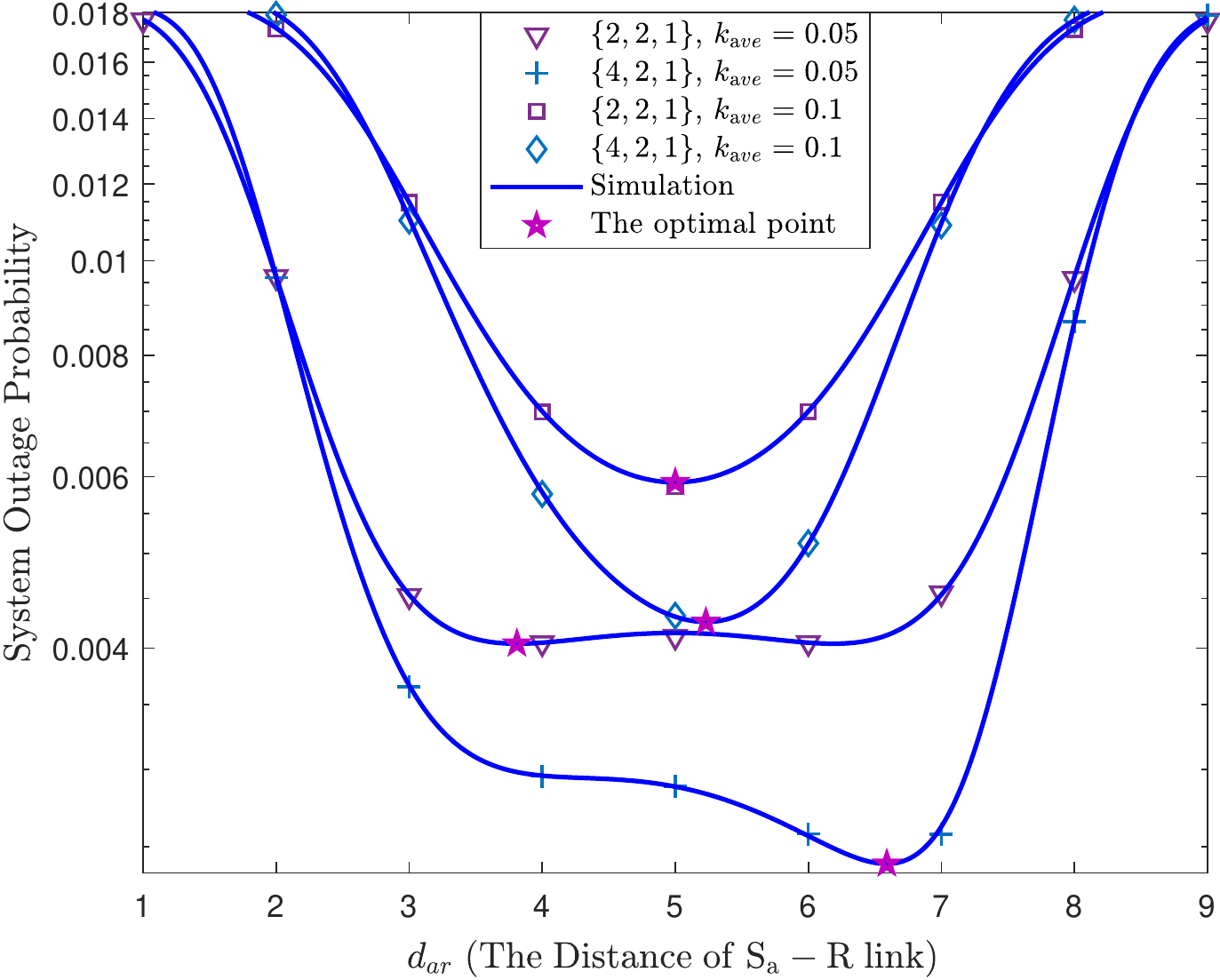}\\
  \caption{$\mathbb{P}_{\rm{out}}$ vs. $d_{ar}$ for various HIs levels and shape parameters, where $\rm R_{th}=0.5$ bit/Hz, $\rho=50$ dB and $d_{br}=10-d_{ar}$.}
\end{figure}

\begin{figure}[!t]
  \centering
  \includegraphics[width=0.45\textwidth]{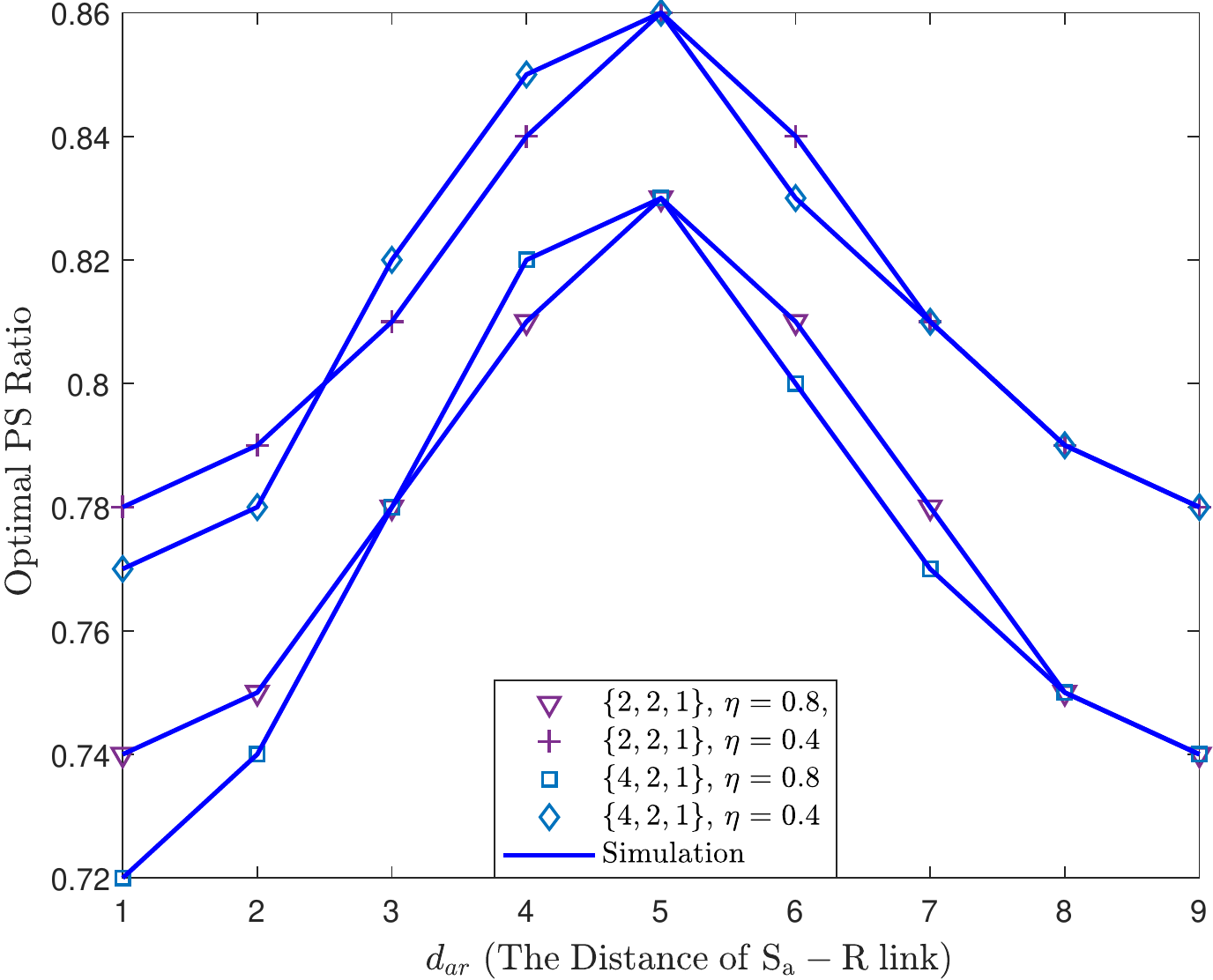}\\
  \caption{Optimal PS ratio vs. $\beta$ for various energy conversion efficiency and shape parameters, where $k_{ave}=0.1$ and $d_{br}=10-d_{ar}$.}
\end{figure}

\begin{figure}[!t]
  \centering
  \includegraphics[width=0.45\textwidth]{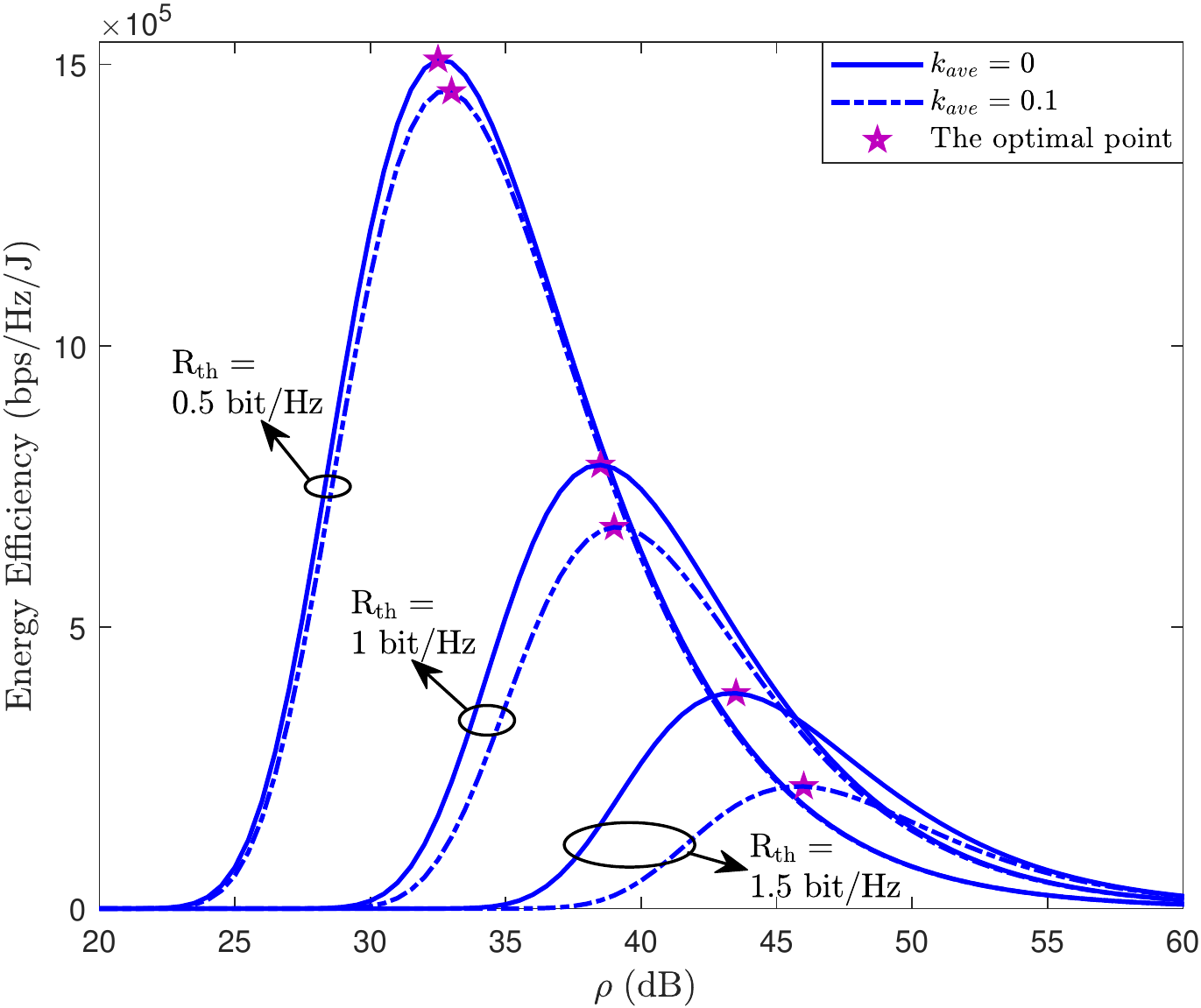}\\ \vspace{-11pt}
  \caption{Energy efficiency vs. $\rho$ for various data rates, where $\{m_a, m_b, m_d\}=\{2, 2, 1\}$.}
\end{figure}

Fig. 10 studies the impacts of relay location on  the system outage probability.
From this figure, we can see that the optimal relay location varies with respect to the channel quality of two relaying links, i.e., $S_a-R$ and $S_b-R$ links. Specifically, for  a fixed $k_{ave}$, the optimal relay location is close to the terminal, where its corresponding relaying link has a smaller shape parameter. This is beneficial for the relay to harvest more energy from both terminals and increase its transmit power. We can also see that  the optimal relay location is closer to the middle as the increase of $k_{ave}$. The above two observations  provide guidances on how to place relay node in the system design.

Fig. 11 plots the optimal PS ratio that minimises the system outage probability versus the relay location. We can see that the optimal PS ratio first increases and then decreases after reaching the maximum value. When $d_{ar}=5$ m, i.e., the relay locates at the middle of the two terminals, the optimal PS ratio reaches the maximum. This indicates that a larger proportion of the received power at the relay is used for EH due to the poorer channel quality of $S_a-R$ and $S_b-R$ links. In addition, for the given shape parameter, the optimal PS ratio decreases as the energy conversion efficiency $\eta$ increases. This is because the relay with a higher $\eta$ can convert more energy from EH, which in turn leaves more of the received power for IP.

Fig. 12 plots the energy efficiency against the input SNR with different data rates and HIs levels. The energy efficiency is computed as $EE = \frac{{T\left( {1 - {\mathbb{P}_{\rm out}}} \right){\rm R_{th}}}}{{2T{P_o}/3}} = \frac{{3\left( {1 - {\mathbb{P}_{\rm out}}} \right){\rm R_{th}}}}{{2{P_o}}}$ \cite{8851300}. One can see that as the increase of SNR, the energy efficiency for all curves firstly increases and then decreases. This indicates that  an optimal SNR exists to maximize the energy efficiency, and   that the selection of appropriate transmit power at terminals is crucial to balance between spectral efficiency and energy efficiency. From this figure, we can also observe that, for a given data rate, the optimal energy efficiency decreases as $k_{ave}$ increases. This is duo to the detrimental impact caused by HIs on system outage performance is more serious with the increase of $k_{ave}$, which further degrades the energy efficiency. Similarly, for a fixed $k_{ave}$, the optimal energy efficiency decreases with data rate increases. 

We summarize the main observations from the simulation results as follows. On the one hand, HIs cause two ceiling effects, i.e., RCC and OSC, which degrade the achievable system outage probability of the considered system. In particular, when  $\gamma_{th}$ exceeds the OSC, the considered system will be in outage and the diversity gain reduces to zero; when  $\gamma_{th}$ lies between the RCC and the OSC, the relaying link will be blocked, in which case the system outage probability is only determined by the direct link and the diversity gain is $m_d$; when $\gamma_{th}$ falls below the RCC, both the relaying link and the direct link will contribute to reducing the system outage probability but the diversity gain is still determined only by the direct link, i.e., $d=m_d$. On the other hand, the system parameters have a big impact on the optimal PS ratio and the desirable relay location. Specifically, the optimal PS ratio that minimises the system outage probability increases as the shape parameters of relaying links increase, as well as the relay approaches the middle between the two terminals; The system outage probability decreases as the relay locates closer to the terminal that has a smaller shape parameter of its relaying link.

\section{Conclusion}
In this paper, we have investigated the system outage performance of a PS-SWIPT based two-way AF relay network with TDBC under Nakagami-$m$ fading channels, while considering the hardware impairments at transceivers. We derived a closed-form expression for the system outage probability as a function of HIs. Based on the derived expression, we  determined two ceiling effects, viz., RCC and OSC, and derived the achievable diversity gain for our considered network.
Besides, our analytical and simulation results have revealed the impact of various system parameters on the system outage performance and the optimal PS ratio, as well as the desirable relay location. The key insights have been summarized in the last paragraph of Section IV.
Finally, we have also provided insights on the adjustment of  transmit power at  terminals to maximize the energy efficiency.


\appendices
\begin{appendices}
\section{}
 According to the expression for $\mathbb{P}_1$, we derive its closed-form expression in the following two cases.

\emph{Case 1:} When ${\gamma _{th}} \ge \frac{1}{{k_1^2 + k_2^2}}$, ${\left( {1 - \left( {k_1^2 + k_2^2} \right){\gamma _{th}}} \right)\rho Z < {\gamma _{th}}}$ is always satisfied no matter what the value of $Z$ ($Z>0$) is. Thus, the outage probability, $\mathbb{P}_1$, is equal to 1.

\emph{Case 2:} When ${\gamma _{th}} < \frac{1}{{k_1^2 + k_2^2}}$, $\mathbb{P}_1$ can be rewritten as $\mathbb{P}_1^1$, given by
\setcounter{equation}{29}
\begin{align}\label{34}
  {\mathbb{P}_1^1} &= \Pr \left( {Z < \frac{{{\gamma _{th}}}}{\rho{\left( {1 - {\gamma _{th}}\left( {k_1^2 + k_2^2} \right)} \right)}}} \right)\nonumber\\
  &= \int_0^{\frac{{{\gamma _{th}}}}{\rho{\left( {1 - {\gamma _{th}}\left( {k_1^2 + k_2^2} \right)} \right)}}} {\frac{1}{{\Gamma \left( {{m_d}} \right)\theta _d^{{m_d}}}}} {z^{{m_d} - 1}}{e^{ - \frac{z}{{{\theta _d}}}}}dz \tag{A.1}.
\end{align}
where ${\theta _d} = \frac{{{\Omega _d}}}{{{m_d}}}$. Using [34, eq.(3.381.1)] and [34, eq.(8.352.6)] to solve the  integration of \eqref{34}, we can obtain the closed-form expression for ${\mathbb{P}_{1}^1}$ as shown in \eqref{17}.

\section{}
Similar to the analysis of $\mathbb{P}_1$, the derivation of $\mathbb{P}_2$ can also be divided into the following two cases.

\emph{Case 1:} When ${{\gamma _{th}} \ge \frac{{{I_1}}}{{{I_2}}}}$, $\left( {{I_1} - {\gamma _{th}}{I_2}} \right)\rho Y< \rho {\gamma _{th}}{I_2}X + {I_3}{\gamma _{th}} + \frac{{{\gamma _{th}}}}{X}$ holds no matter what the value of $X$ ($X>0$) is. Thus, the outage probability, $\mathbb{P}_2$, is equal to 1.

\emph{Case 2:} When ${{\gamma _{th}} < \frac{{{I_1}}}{{{I_2}}}}$ is satisfied, $\mathbb{P}_2$ can be re-expressed as $\mathbb{P}_2^1$, given by
\begin{align}\label{51}
&{\mathbb{P}_2^1} = \Pr \left( {Y < {I_4}{\gamma _{th}}\left( {\rho {I_2}X + {I_3} + \frac{1}{X}} \right)} \right)\nonumber\\
&= \int_0^\infty  {\int_0^{Q(x)} {{f_X}\left( x \right){f_Y}\left( y \right)dydx} }\nonumber\\
&= \frac{1}{{\Gamma \left( {{m_a}} \right)\Gamma \left( {{m_b}} \right)\theta _a^{{m_a}}}}\nonumber\\
&\times \int_0^\infty  {{x^{{m_a} - 1}}{e^{ - \frac{x}{{{\theta _a}}}}}} \gamma \left( {{m_b},\frac{{{I_4}{\gamma _{th}}}}{{{\theta _b}}}\left( {\rho {I_2}x + {I_3} + \frac{1}{x}} \right)} \right)dx,\tag{B.1}
\end{align}
where ${\theta _a} = \frac{{{\Omega _a}}}{{{m_a}}}$, ${\theta _b} = \frac{{{\Omega _b}}}{{{m_b}}}$, $I_4=\frac{1}{{\left( {{I_1} - {\gamma _{th}}{I_2}} \right)\rho }}$ and $Q(x)={I_4}{\gamma _{th}}\left( {\rho {I_2}x + {I_3} + \frac{1}{x}} \right)$.

Using [34, eq.(3.381.1)] and polynomial expansion for ${\left( {\rho {I_2}x + {I_3} + \frac{1}{x}} \right)}^l$, $\mathbb{P}_2^1$ can be given as
\begin{align}\label{19}
{\mathbb{P}_2^1} = &1 - \frac{1}{{\Gamma \left( {{m_a}} \right)\theta _a^{{m_a}}}}{e^{ - \frac{{{I_3}{I_4}{\gamma _{th}}}}{{{\theta _b}}}}}\sum\limits_{l = 0}^{{m_b} - 1} {\sum\limits_{s = 0}^l {\sum\limits_{t = 0}^{l - s} {\left( \begin{array}{l}
l\\
s
\end{array} \right)} } }\nonumber\\
& \times \left( \begin{array}{l}
l - s\\
t
\end{array} \right)\frac{1}{{l!}}{\left( {{I_2}\rho } \right)^s}I_3^{l - s - t}{\left( {\frac{{{I_4}{\gamma _{th}}}}{{{\theta _b}}}} \right)^l}\nonumber\\
 &\times \int_0^\infty  {{x^{s + {m_a} - t - 1}}{e^{ - \left( {\frac{{{I_2}{I_4}\rho {\gamma _{th}}}}{{{\theta _b}}} + \frac{1}{{{\theta _a}}}} \right)x - \frac{{{I_4}{\gamma _{th}}}}{{{\theta _b}x}}}}dx}.\tag{B.2}
\end{align}
Utilizing [34, eq.(3.471.9)] to address the integration of \eqref{19}, we can also obtain the closed-form expression for $\mathbb{P}_{2}^1$ as given in \eqref{18}.

\section{}
Referring to the derivation of $\mathbb{P}_1$ and $\mathbb{P}_2$, we can also derive the closed-form expression for $\mathbb{P}_4$ in the following two cases.

\emph{Case 1:} When ${{\gamma _{th}} \ge \frac{{{I_1}}}{{{I_2}}}}$, both inequalities of $\mathbb{P}_4$ hold regardless of the value of $X$ and $Y$. Thus, the outage probability, $\mathbb{P}_4$, is equal to 1.

\emph{Case 2:} When ${{\gamma _{th}} < \frac{{{I_1}}}{{{I_2}}}}$, the $\mathbb{P}_4$ can be re-expressed as ${\mathbb{P}_4} = \Pr \left( {Y < {\Delta _a},X < {\Delta _b}} \right)$, where ${\Delta _a} = {I_4}{\gamma _{th}}\left( {\rho {I_2}X + {I_3} + \frac{1}{X}} \right)$ and ${\Delta _b} = {I_4}{\gamma _{th}}\left( {\rho {I_2}Y + {I_3} + \frac{1}{Y}} \right)$. Due to the high correlation between ${\Delta _a}$ and ${\Delta _b}$, it is hard to derive $\mathbb{P}_4$ directly. Therefore, we first analyze and determine the integral region. Based on the integral region, we further obtain the close-form expression for $\mathbb{P}_4$ by calculating the integral value.

Let $X=x$ and $Y=y$. From the expression of $\mathbb{P}_4$ in \eqref{15}, the integral region of $\mathbb{P}_4$ is bounded by two line, which are ${l_1}:y = {I_4}{\gamma _{th}}\left( {\rho {I_2}x + {I_3} + \frac{1}{x}} \right)$ and ${l_2}:x = {I_4}{\gamma _{th}}\left( {\rho {I_2}y + {I_3} + \frac{1}{y}} \right)$. In addition, we define ${l_3}$ as $y = x$. Through observation and some mathematical calculations for the above three curves, we draw the following conclusions:

1) the curves $l_1$ and $l_2$ are symmetric about $l_3$;

2) for the curve $l_1$ (or $l_2$), $y$ (or $x$)  decreases monotonically when $x$ (or $y$) increases from 0 to $\frac{1}{{\sqrt {\rho {I_2}} }}$, whereas increases monotonically when $x$ (or $y$) grows from $\frac{1}{{\sqrt {\rho {I_2}} }}$ to $\infty $;

3) $x = \frac{1}{{\sqrt {\rho {I_2}} }}$ (or $y = \frac{1}{{\sqrt {\rho {I_2}} }}$) is the unique minimum of curve $l_1$ (or $l_2$);

4) $\mathop {\lim }\limits_{x \to 0/\infty} \left( {{I_4}{\gamma _{th}}\left( {\rho {I_2}x + {I_3} + \frac{1}{x}} \right)} \right) = \infty $ and $\mathop {\lim }\limits_{y \to 0/ \infty } \left( {{I_4}{\gamma _{th}}\left( {\rho {I_2}y + {I_3} + \frac{1}{y}} \right)} \right) = \infty $;

5) when $\frac{{{I_1}}}{{2{I_2}}} \le {\gamma _{th}} < \frac{{{I_1}}}{{{I_2}}}$, there is no intersection between curve $l_1$ and curve $l_3$ in the first quadrant;

6) when ${\gamma _{th}} < \frac{{{I_1}}}{{2{I_2}}}$, curve $l_1$ and curve $l_3$ have only one intersection $x_{in}$ in the first quadrant, and ${x_{in}} = \frac{{ - {I_3}{\gamma _{th}} - \sqrt {{{\left( {{I_3}{\gamma _{th}}} \right)}^2} - 4\rho {\gamma _{th}}\left( {2{I_2}{\gamma _{th}} - {I_1}} \right)} }}{{2\rho \left( {2{I_2}{\gamma _{th}} - {I_1}} \right)}}$.
\begin{figure}[!t]
  \centering
  \includegraphics[width=0.28\textwidth]{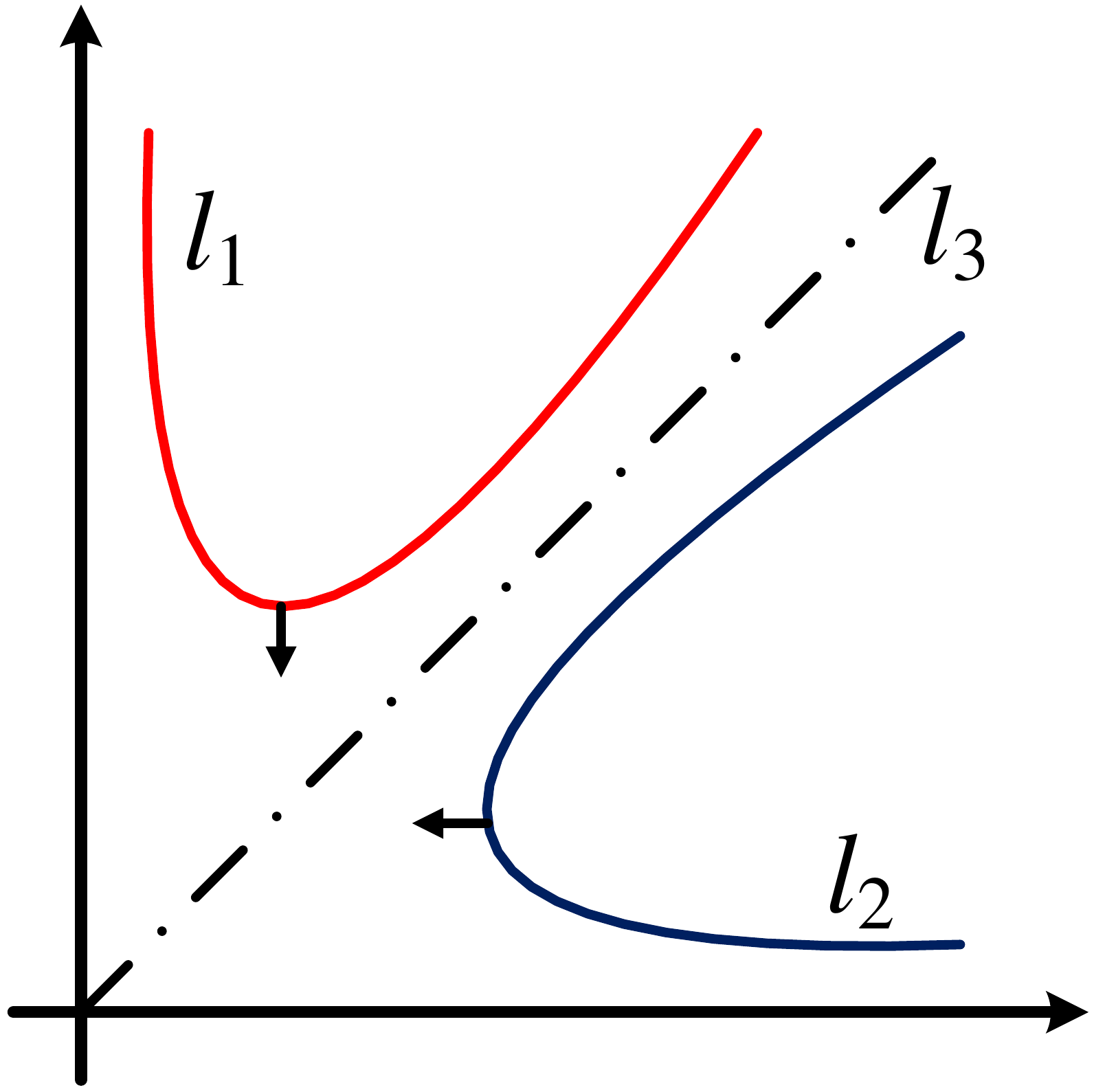} \\
  \caption{The integral region for $\mathbb{P}_4$, where $l_1$ and $l_2$ have no intersection point.}\vspace{-15pt}
\end{figure}

Based on the number of intersection points between curve $l_1$ and curve $l_3 $, there can be two cases for $\mathbb{P}_4$, discussed as follows.

\emph{i):} When the curves $l_1$ and $l_3$ have no intersection point, viz., $\frac{{{I_1}}}{{2{I_2}}} \le {\gamma _{th}} < \frac{{{I_1}}}{{{I_2}}}$, there is also no intersection point between curve $l_1$ and curve $l_2$ in the first quadrant. In this case, the integral region for $\mathbb{P}_4$ is shown in Fig. 13 and $\mathbb{P}_4$ can be written as $\mathbb{P}_4^1$, given by
\setcounter{equation}{32}
\begin{align}
  {\mathbb{P}_4^1} =& 1 - \int_0^\infty  {\int_{Q(x)}^\infty  {{f_X}\left( x \right){f_Y}\left( y \right)dydx} }\nonumber\\
&- \int_0^\infty  {\int_{Q(y)}^\infty  {{f_Y}\left( y \right){f_X}\left( x \right)dxdy} } \tag{C.1}.
\end{align}
Similar to the derivation of $\mathbb{P}^1_{2}$ and $\mathbb{P}^1_{3}$, $\mathbb{P}_{4}^1$ can be calculated as \eqref{22}.

\emph{ii):} When the curves $l_1$ and $l_3$ have only one intersection point, viz., ${\gamma _{th}} < \frac{{{I_1}}}{2{{I_2}}}$, there exist one or three intersection point(s) between curve $l_1$ and curve $l_2$ in the first quadrant. In this case, it is difficult to determine the integral region for $P_4$ directly.The intersection points can be obtained by solving the following equations, given by
\setcounter{equation}{33}
\begin{align}\label{23}
  \left\{ \begin{array}{l}
y = {I_4}{\gamma _{th}}( {\rho {I_2}x + {I_3} + \frac{1}{x}} )\\
x = {I_4}{\gamma _{th}}( {\rho {I_2}y + {I_3} + \frac{1}{y}} )
\end{array} \right..\tag{C.2}
\end{align}
Through some mathematical manipulation, \eqref{23} can be re-expressed as the following quartic function, given as
\begin{align}\label{24}
{c_4}{x^4} + {c_3}{x^3} + {c_2}{x^2} + {c_1}x + {c_0} = 0,\tag{C.3}
\end{align}
where
\begin{align}\notag
{c_0} &= \rho {I_2}I_4^2\gamma _{th}^3,\;{c_1} = {I_3}{I_4}\gamma _{th}^2\left( {1 + 2\rho {\gamma _{th}}{I_2}{I_4}} \right),\\ \notag
{c_2} &= {I_4}\gamma _{th}^2\left( {\rho {\gamma _{th}}{I_2}{I_4}I_3^2 + 2{\gamma _{th}}{I_4}{\rho ^2}I_2^2 + I_3^2}\right),\\ \notag
{c_3} &= {I_3}{\gamma _{th}}\left( {2{\rho ^2}\gamma _{th}^2I_2^2I_4^2 + \rho {\gamma _{th}}{I_2}{I_4} - 1} \right), \\ \notag
{c_4} &= \rho {\gamma _{th}}{I_2}\left( {{\rho ^2}\gamma _{th}^2I_2^2I_4^2 - 1} \right).
\end{align}

\begin{figure}[!t]
  \centering
  \includegraphics[width=0.29\textwidth]{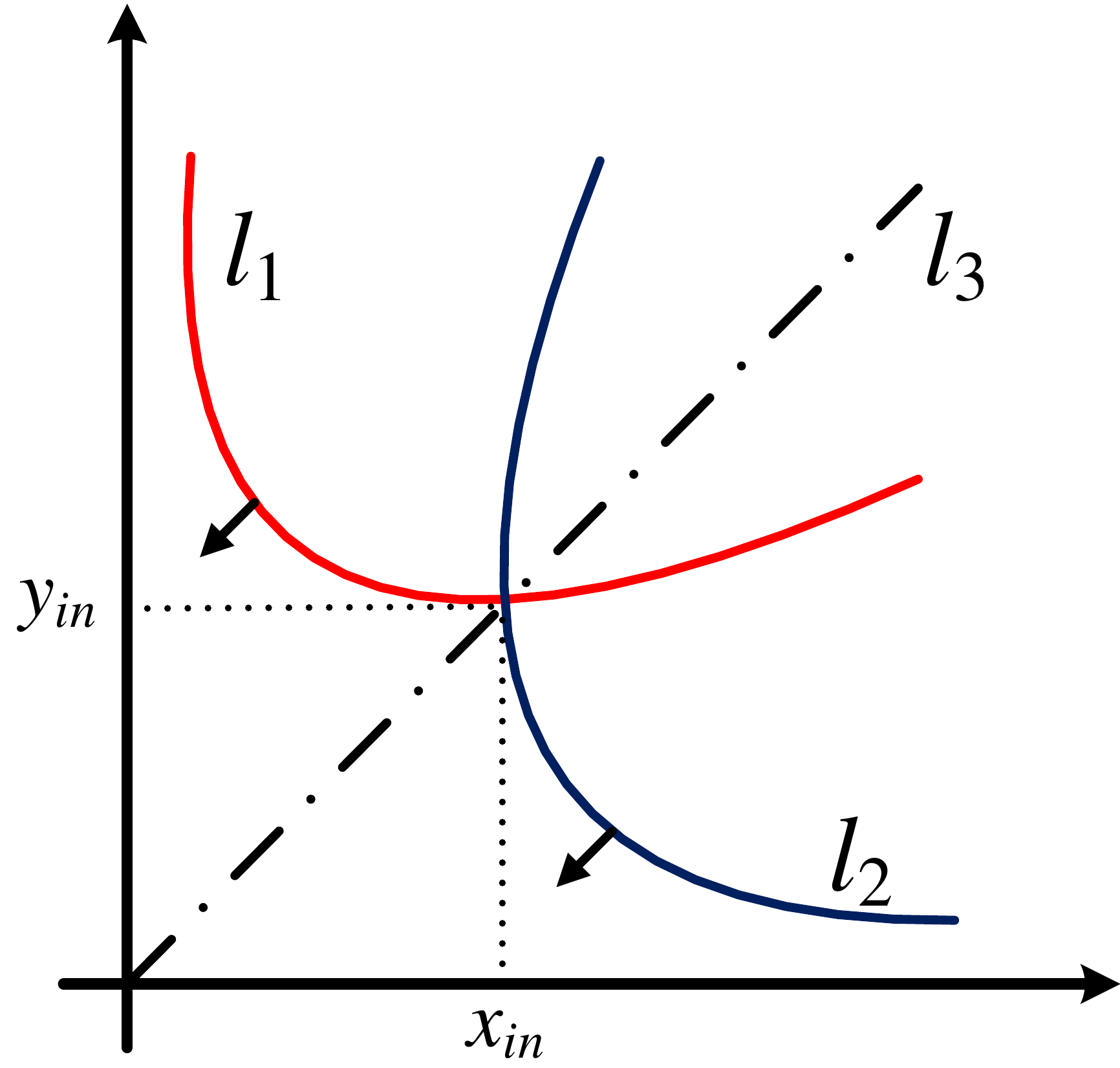} \\
  \caption{The integral region for $\mathbb{P}_4$, where $l_1$ and $l_2$ have only one intersection point.}\vspace{-15pt}
\end{figure}

Obviously, the quartic function \eqref{24}  has at most four positive real roots. Combined with the above conclusions, we can determine that \eqref{24} has one positive real ($x_{in}$) root or three different positive real roots ($x_1$, $x_2$ and $x_{in}$ ), where the closed-form real root(s) can be easily obtained \cite{QuarticfunctionHandbook}. In other words, curves $l_1$ and $l_2$ has one or three different intersection points in the first quadrant. According to the number of positive real root for \eqref{24}, there can be two scenarios for $\mathbb{P}_4$, discussed as follows.

\emph{1):} When the quartic function \eqref{24} have only one positive real root $x_{in}$, the integral region for $\mathbb{P}_4$ can be shown in Fig. 14. Thus, $\mathbb{P}_4$ can be computed as $\mathbb{P}_4^2$, given by
\begin{align}\label{26}
{\mathbb{P}_4^2} =& 1 - \Bigg(\underbrace {\int_0^{{x_{in}}} {\int_{Q(x)}^\infty  {{f_X}\left( x \right){f_Y}\left( y \right)dydx} } }_{\mathbb{P}_{4\_1}^2}\nonumber\\
& + \underbrace {\int_{{x_{in}}}^\infty  {\int_x^\infty  {{f_X}\left( x \right){f_Y}\left( y \right)dydx} } }_{\mathbb{P}_{4\_2}^2}\nonumber\\
& + \underbrace {\int_0^{{x_{in}}} {\int_{Q(y)}^\infty  {{f_Y}\left( y \right){f_X}\left( x \right)dxdy} } }_{\mathbb{P}_{4\_3}^2}\nonumber\\
&+ \underbrace {\int_{{x_{in}}}^\infty  {\int_y^\infty  {{f_Y}\left( y \right){f_X}\left( x \right)dxdy} } }_{\mathbb{P}_{4\_4}^2}\Bigg).\tag{C.4}
\end{align}

\begin{figure*}[!t]
\normalsize
\setcounter{equation}{24}

 \begin{align}\label{29}
 &\mathbb{P}_{4\_1}^3 + \mathbb{P}_{4\_2}^3 = \underbrace {\frac{1}{{\Gamma \left( {{m_a}} \right)\theta _a^{{m_a}}}}{e^{ - \frac{{{\gamma _{th}}{I_3}{I_4}}}{{{\theta _b}}}}}\sum\limits_{l = 0}^{{m_b} - 1} {\frac{1}{{l!}}{{\left( {\frac{{{I_4}{\gamma _{th}}}}{{{\theta _b}}}} \right)}^l}} \int_{{x_{in}}}^{{\Phi _1}} {{x^{{m_a} - 1}}{{\left( {\rho {I_2}x + {I_3} + \frac{1}{x}} \right)}^l}{e^{ - \left( {\frac{{\rho {\gamma _{th}}{I_2}{I_4}}}{{{\theta _b}}} + \frac{1}{{{\theta _a}}}} \right)x - \frac{{{\gamma _{th}}{I_4}}}{{{\theta _b}x}}}}dx} }_{{\Xi _3}}\nonumber\\
 &+ \underbrace {\frac{1}{{\Gamma \left( {{m_a}} \right)\theta _a^{{m_a}}}}\sum\limits_{l = 0}^{{m_b} - 1} {\frac{1}{{l!}}{{\left( {\frac{1}{{{\theta _b}}}} \right)}^l}} \int_{{\Phi _1}}^\infty  {{x^{{m_a} - 1}}{G^l}\left( x \right){e^{ - \left( {\frac{{G\left( x \right)}}{{{\theta _b}}} + \frac{x}{{{\theta _a}}}} \right)}}} dx}_{{\Xi _4}} - \underbrace {\frac{1}{{\Gamma \left( {{m_a}} \right)\theta _a^{{m_a}}}}\sum\limits_{l = 0}^{{m_b} - 1} {\frac{1}{{l!}}{{\left( {\frac{1}{{{\theta _b}}}} \right)}^l}} \int_{{x_{in}}}^\infty  {{x^{l + {m_1} - 1}}{e^{ - \left( {\frac{{{\theta _a} + {\theta _b}}}{{{\theta _a}{\theta _b}}}} \right)x}}} dx}_{{\Xi _5}}.\tag{C.11}
\end{align}
\setcounter{equation}{22}
\hrulefill
\end{figure*}

According to \eqref{2} and \eqref{25}, $\mathbb{P}_{4\_1}^2 + \mathbb{P}_{4\_2}^2$ can be calculated as
\begin{align}\label{27}
  \mathbb{P}_{4\_1}^2 + \mathbb{P}_{4\_2}^2&= 1- \underbrace {\int_{{x_{in}}}^\infty  {{f_X}\left( x \right){F_Y}\left( x \right)dx} }_{{\Xi _1}}\nonumber \\
  & - \underbrace {\int_0^{{x_{in}}} {{f_X}\left( x \right){F_Y}\left( {Q(x)} \right)dx} }_{{\Xi _2}}.\tag{C.5}
\end{align}

Based on [34, eq.(3.381.1)], the first term of \eqref{27}, ${{\Xi _1}}$, can be written as
\begin{align}
{\Xi _1} =& \int_{{x_{in}}}^\infty  {{f_X}\left( x \right)dx}  - \frac{1}{{\Gamma \left( {{m_a}} \right)\theta _a^{{m_a}}}}\sum\limits_{l = 0}^{{m_b} - 1} {\frac{1}{{l!}}{{\left( {\frac{1}{{{\theta _b}}}} \right)}^l}}\nonumber\\
&\times \int_{{x_{in}}}^\infty  {{x^{l + {m_a} - 1}}{e^{ - \left( {\frac{1}{{{\theta _a}}} + \frac{1}{{{\theta _b}}}} \right)x}}} dx.\tag{C.6}
\end{align}
Using \eqref{25} and [34, eq.(3.381.3)], ${\Xi _1}$ can be calculated as
\begin{align}
{\Xi _1} =& 1\! -\!  \frac{1}{{\Gamma \left( {{m_a}} \right)}}\gamma \left( {{m_a},\frac{{{x_{in}}}}{{{\theta _a}}}} \right)  - \frac{1}{{\Gamma \left( {{m_a}} \right)\theta _a^{{m_a}}}}\sum\limits_{l = 0}^{{m_b} - 1} {\frac{1}{{l!}}} {\left( {\frac{1}{{{\theta _b}}}} \right)^l}\nonumber\\
 & \times {\left( {\frac{1}{{{\theta _a}}} + \frac{1}{{{\theta _b}}}} \right)^{ - \left( {l + {m_a}} \right)}}\Gamma \left( {l + {m_a},{x_{in}}\left( {\frac{1}{{{\theta _a}}} + \frac{1}{{{\theta _b}}}} \right)} \right).\tag{C.7}
\end{align}

The second term of \eqref{27}, ${{\Xi _2}}$, can be written, using [34, eq.(3.381.1)], as
\begin{align}\label{28}
{\Xi _2} = &\frac{1}{{\Gamma \left( {{m_a}} \right)}}\gamma \left( {{m_a},\frac{{{x_{in}}}}{{{\theta _a}}}} \right) - \frac{1}{{\Gamma \left( {{m_a}} \right)\theta _a^{{m_a}}}}{e^{ - \frac{{{\gamma _{th}}{I_3}{I_4}}}{{{\theta _b}}}}}\nonumber\\
 &\times {\sum\limits_{l = 0}^{{m_b} - 1} {\frac{1}{{l!}}\left( {\frac{{{I_4}{\gamma _{th}}}}{{{\theta _b}}}} \right)} ^l}\int_0^{{x_{in}}} {{x^{{m_a} - 1}}{{\left( {\rho {I_2}x + {I_3} + \frac{1}{x}} \right)}^l}}\nonumber\\
 &\times {e^{ - \left( {\frac{{\rho {\gamma _{th}}{I_2}{I_4}}}{{{\theta _b}}} + \frac{1}{{{\theta _a}}}} \right)x - \frac{{{\gamma _{th}}{I_4}}}{{{\theta _b}x}}}}dx.\tag{C.8}
\end{align}
Since the closed-form expression for the required integral in \eqref{28} can not be obtained directly, we adopt the Guassian-Chebyshev quadrature\footnote{In this paper, we adopt the Gaussian-Chebyshev quadrature instead of other approximation methods because it can provide sufficient level of accuracy with very few terms. Thanks to this advantage, the Gaussian-Chebyshev quadrature has been widely used in the state-of-the-art works \cite{8613860,8633928,8576644}.} to approximate it. Thus, ${\Xi _2}$ can be approximated as
\begin{align}\label{30}
  {\Xi _2} &= \frac{1}{{\Gamma \left( {{m_a}} \right)}}\gamma \left( {{m_a},\frac{{{x_{in}}}}{{{\theta _a}}}} \right) - \frac{{\pi {x_{in}}}}{{2N\Gamma \left( {{m_a}} \right)\theta _a^{{m_a}}}}{e^{ - \frac{{{\gamma _{th}}{I_3}{I_4}}}{{{\theta _b}}}}}\nonumber\\
   &\times {\sum\limits_{l = 0}^{{m_b} - 1} {\sum\limits_{n = 1}^N {\frac{1}{{l!}}\left( {\frac{{{I_4}{\gamma _{th}}}}{{{\theta _b}}}} \right)} } ^l}\sqrt {1 - v_n^2} {\left( {k_n^1} \right)^{{m_a} - 1}}\nonumber\\
    &\times {\left( {\rho {I_2}k_n^1 + {I_3} + \frac{1}{{k_n^1}}} \right)^l}{e^{ - \left( {\frac{{\rho {\gamma _{th}}{I_2}{I_4}}}{{{\theta _b}}} + \frac{1}{{{\theta _a}}}} \right)k_n^1 - \frac{{{\gamma _{th}}{I_4}}}{{{\theta _b}k_n^1}}}},\tag{C.9}
\end{align}
where ${v_n} = \cos \left( {\frac{{2n - 1}}{{2N}}\pi } \right)$, $k_n^1 = \frac{{{x_{in}}}}{2}{v_n} + \frac{{{x_{in}}}}{2}$ and $N$ is the parameter that determines the tradeoff between complexity and accuracy.

Referring to the derivation of $\mathbb{P}_{4\_1}^2 + \mathbb{P}_{4\_2}^2$, we can also obtain the closed-form expression for $\mathbb{P}_{4\_3}^2 + \mathbb{P}_{4\_4}^2$. With this, we can obtain the closed-form expression for $\mathbb{P}^2_4$, as shown in \eqref{29a}.

\begin{figure}[!t]
  \centering
  \includegraphics[width=0.28\textwidth]{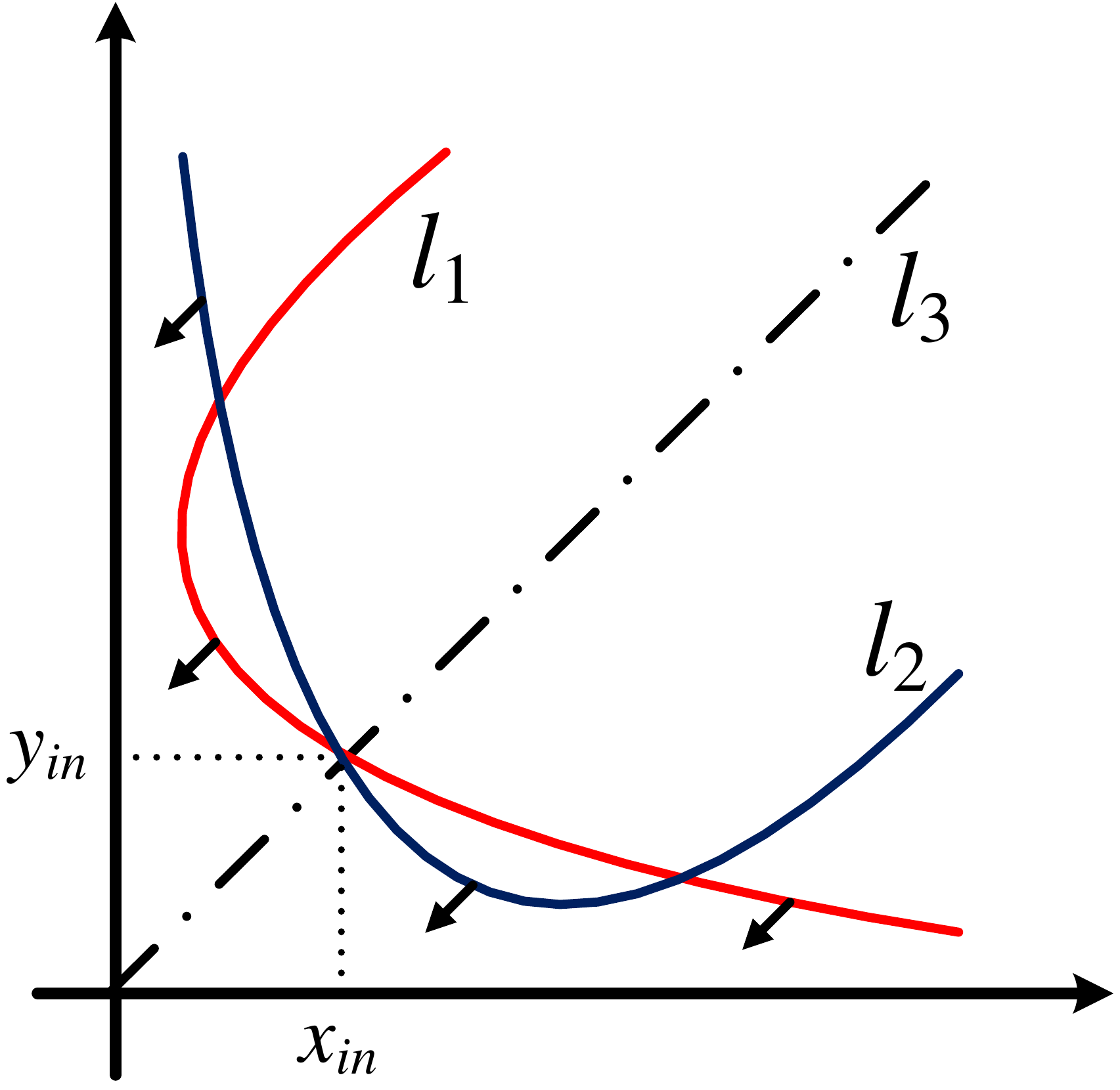}\\ \vspace{-5pt}
  \caption{The integral region for $\mathbb{P}_4$, where $l_1$ and $l_2$ have three intersection points.}\vspace{-15pt}
\end{figure}

\emph{2):} When the quartic function \eqref{24} have three three different positive real roots ($x_1$, $x_2$ and $x_{in}$), the integral region for $\mathbb{P}_4$ can be shown in Fig. 15. Obviously, $\mathbb{P}_4$ can be calculated as ${\mathbb{P}_4^3}$, given by
\setcounter{equation}{41}
\begin{align}
{\mathbb{P}_4^3} =& U\left( {{K_1} - {K_2}} \right)\nonumber\\
&\times \Bigg(1 - \Bigg(\underbrace {\int_{{x_{in}}}^{{\Phi _1}} {\int_{Q(x)}^x {{f_X}\left( x \right){f_Y}\left( y \right)dydx} } }_{\mathbb{P}_{4\_1}^3}\nonumber\\
&+ \underbrace {\int_{{\Phi _1}}^\infty  {\int_{G\left( x \right)}^x {{f_X}\left( x \right){f_Y}\left( y \right)dydx} } }_{\mathbb{P}_{4\_2}^3}\nonumber\\
 &+ \underbrace {\int_{{x_{in}}}^{{\Phi _2}} {\int_{Q(y)}^y {{f_Y}\left( y \right){f_X}\left( x \right)dxdy} } }_{\mathbb{P}_{4\_3}^3}\nonumber\\
&+ \underbrace {\int_{{\Phi _2}}^\infty  {\int_{G\left( y \right)}^y {{f_Y}\left( y \right){f_X}\left( x \right)dxdy} } }_{\mathbb{P}_{4\_4}^3}\Bigg)\Bigg)\nonumber\\
&+ (1 - U\left( {{K_1} - {K_2}} \right))\mathbb{P}_4^2,\tag{C.10}
\end{align}
with
\begin{align}\notag
&U\left( x \right) = \left\{ \begin{array}{l}
1,\;\;x > 0,\\
0,\;\;x \le 0,\\
\end{array} \right. \notag
 \end{align}
 \begin{align}\notag
 &{\Phi _1} = \max \left\{ {{x_1},{x_2}} \right\}, \; {\Phi _2} = \max \left\{ {Q(x_1),Q(x_2)} \right\},\\ \notag
 &G\left( x \right) = \frac{{ - \left( {{\gamma _{th}}{I_3}{I_4} - x} \right) - \sqrt {{{\left( {{\gamma _{th}}{I_3}{I_4} - x} \right)}^2} - 4\rho \gamma _{th}^2{I_2}I_4^2} }}{{2\rho {\gamma _{th}}{I_2}{I_4}}},\\ \notag
 &{K_1} = G\left( {\frac{{{x_{in}} + {\Phi _1}}}{2}} \right),{K_2} = Q\left( {\frac{{{x_{in}} + {\Phi _1}}}{2}} \right).
 \end{align}
Based on \eqref{2}, \eqref{25}, and some mathematical calculations, $\mathbb{P}_{4\_1}^3 + \mathbb{P}_{4\_2}^3$ can be expressed as \eqref{29} at the top of previous page. Similar to \eqref{30}, the first term of \eqref{29}, ${\Xi _3}$, can be approximated, using the Gaussian-Chebyshev quadrature, as
\begin{align}
{\Xi _3} = &\frac{{\pi \left( {{\Phi _1} - {x_{in}}} \right)}}{{2N\Gamma \left( {{m_a}} \right)\theta _a^{{m_a}}}}{e^{ - \frac{{{\gamma _{th}}{I_3}{I_4}}}{{{\theta _b}}}}}\sum\limits_{l = 0}^{{m_b} - 1} {\sum\limits_{n = 1}^N {\frac{1}{{l!}}{{\left( {\frac{{{I_4}{\gamma _{th}}}}{{{\theta _b}}}} \right)}^l}} }\nonumber\\
& \times \sqrt {1 - v_n^2} {\left( {k_n^2} \right)^{{m_a} - 1}}{\left( {\rho {I_2}k_n^2 + {I_3} + \frac{1}{{k_n^2}}} \right)^l}\nonumber\\
&\times {e^{ - \left( {\frac{{\rho {\gamma _{th}}{I_2}{I_4}}}{{{\theta _b}}} + \frac{1}{{{\theta _a}}}} \right)k_n^2 - \frac{{{\gamma _{th}}{I_4}}}{{{\theta _b}k_n^2}}}},\tag{C.12}
\end{align}
where $k_n^2 = \frac{{{\Phi _1} - {x_{in}}}}{2}{v_n} + \frac{{{\Phi _1} + {x_{in}}}}{2}$.

Utilizing the variable substitution $y = \tan \theta $ and the Gaussian-Chebyshev quadrature, the second term, ${\Xi _4}$, can be calculated as
\begin{align}
{\Xi _4} = &\frac{{{\pi ^2}}}{{4N\Gamma \left( {{m_a}} \right)\theta _a^{{m_a}}}}\sum\limits_{l = 0}^{{m_b} - 1} {\sum\limits_{n = 1}^N {\frac{1}{{l!}}{{\left( {\frac{1}{{{\theta _b}}}} \right)}^l}} }\nonumber\\
&\times \sqrt {1 - v_n^2} {\left( {\tan k_n^3 + {\Phi _1}} \right)^{{m_a} - 1}}{G^l}\left( {\tan k_n^3 + {\Phi _1}} \right)\nonumber\\
&\times {\sec ^2}\left( {k_n^3} \right){e^{ - \left( {\frac{{G\left( {\tan k_n^3 + {\Phi _1}} \right)}}{{{\theta _b}}} + \frac{{\tan k_n^3 + {\Phi _1}}}{{{\theta _a}}}} \right)}},\tag{C.13}
\end{align}
where $k_n^3 = \frac{\pi }{4}{v_n} + \frac{\pi }{4}$.

The three term, ${\Xi _5}$, can be given, using [34, eq.(3.381.3)], as
\begin{align}
 {\Xi _5} = &\frac{1}{{\Gamma \left( {{m_a}} \right)\theta _a^{{m_a}}}}\sum\limits_{l = 0}^{{m_b} - 1} {\frac{1}{{l!}}{{\left( {\frac{1}{{{\theta _b}}}} \right)}^l}} {\left( {\frac{{{\theta _a} + {\theta _b}}}{{{\theta _a}{\theta _b}}}} \right)^{ - \left( {l + {m_a}} \right)}}\nonumber\\
 &\times \Gamma \left( {l + {m_a},\frac{{{x_{in}}\left( {{\theta _a} + {\theta _b}} \right)}}{{{\theta _a}{\theta _b}}}} \right).\tag{C.14}
\end{align}
 Similar to the derivation of $\mathbb{P}_{4\_1}^3 + \mathbb{P}_{4\_2}^3$, we can also obtain the closed-form expression for $\mathbb{P}_{4\_3}^3 + \mathbb{P}_{4\_4}^3$. With this, we can obtain the closed-form expression for $\mathbb{P}_4^3$, as shown in \eqref{29b}.

\end{appendices}
\vspace{-5pt}
\nocite{Gradshteyn2007Table}

\ifCLASSOPTIONcaptionsoff
 \newpage
\fi
\bibliographystyle{IEEEtran}
\bibliography{refa}
\end{document}